\documentclass[aps,pra,twocolumn,groupedaddress,nofootinbib]{revtex4-1}
\usepackage{amssymb,amsmath,mleftright,mathtools}
\usepackage{stix}
\usepackage{graphicx}
\usepackage{units}
\usepackage[colorlinks,linkcolor=blue,citecolor=blue,urlcolor=blue]{hyperref}
\usepackage{cleveref}
\usepackage{bm}
\newcommand{\nn} {\nonumber}

\newcommand{\ket}{\rangle}
\newcommand{\bra}{\langle}

\newcommand{\hpsi}{\hat{\psi}}

\newcommand{\mP}{\mathcal{P}}
\newcommand{\mC}{\mathcal{C}}
\newcommand{\mH}{\mathcal{H}}

\newcommand{\w}{\omega}
\newcommand{\dd}{\mathrm{d}}
\newcommand{\ii}{i}

\newcommand{\cp}{\mathscr{p}}
\newcommand{\cq}{\mathscr{q}}
\newcommand{\cpu}{\underline{\mathscr{p}}}
\newcommand{\cqu}{\underline{\mathscr{q}}}

\newcommand{\bvec}[1]{\mathbf{#1}}
\renewcommand{\vec}[1]{\bm{\mathit{#1}}}

\newcommand{\bx}{\mathbf{x}}

\newcommand{\nF}{n_\text{F}}
\newcommand{\aB}{a_\text{B}}
\newcommand{\nbF}{\bar{n}_\text{F}}
\newcommand{\nFs}[1]{n_{\text{F},#1}}

\newcommand{\nB}{n_\text{B}}
\newcommand{\kF}{k_\text{F}}
\newcommand{\vF}{v_\text{F}}
\newcommand{\eF}{\epsilon_\text{F}}

\newcommand{\gR}{g^{\text{R}}}
\newcommand{\SgR}{\Sigma^{\text{R}}}
\newcommand{\SgRc}{\Sigma^{\text{R}}_c}
\newcommand{\WnR}{W_0^{\text{R}}}
\newcommand{\WnT}{W_0^{\text{T}}}
\newcommand{\WT}{W^{\text{T}}}
\newcommand{\WnAT}{W_0^{\overline{\text{T}}}}
\newcommand{\eR}{\varepsilon^{\text{R}}}
\newcommand{\enR}{\varepsilon_0^{\text{R}}}

\DeclareMathOperator{\thf}{\theta}
\DeclareMathOperator{\dlf}{\delta}
\DeclareMathOperator{\sgn}{sign}
\DeclareMathOperator{\im}{Im}
\DeclareMathOperator{\re}{Re}
\DeclarePairedDelimiter\abs{\lvert}{\rvert}

\graphicspath{{./Figs/}}

\begin{document}
\title{Dynamically screened vertex correction to $GW$}
\author{Y. Pavlyukh}
\email[]{yaroslav.pavlyukh@gmail.com}
\affiliation{Institut f\"{u}r Physik, Martin-Luther-Universit\"{a}t Halle-Wittenberg,
  06120 Halle, Germany}
\author{G. Stefanucci}
%\email[]{gianluca.stefanucci@roma2.infn.it}
\affiliation{Dipartimento di Fisica, Universit{\`a} di Roma Tor Vergata, 
Via della Ricerca Scientifica 1, 00133 Rome, Italy}
\author{R. van Leeuwen}
%\email[]{robert.vanleeuwen@jyu.fi}
\affiliation{Department of Physics, Nanoscience Center, University of Jyv{\"a}skyl{\"a}, 
FI-40014 Jyv{\"a}skyl{\"a}, Finland}
\date{\today}
\begin{abstract}
  %-----------------------------------------------------------------------------------------
  Diagrammatic perturbation theory is a powerful tool for the investigation of interacting
  many-body systems, the self-energy operator $\Sigma$ encoding all the variety of
  scattering processes. In the simplest scenario of correlated electrons described by the
  $GW$ approximation for the electron self-energy, a particle transfers a part of its
  energy to neutral excitations.  Higher-order (in screened Coulomb interaction $W$)
  self-energy diagrams lead to improved electron spectral functions (SF) by taking more
  complicated scattering channels into account and by adding corrections to lower order
  self-energy terms. However, they also may lead to unphysical negative spectral
  function. The resolution of this difficulty has been demonstrated in our previous
  works. The main idea is to represent the self-energy operator in a Fermi Golden rule
  form which leads to the manifestly positive definite SF and allows for a very efficient
  numerical algorithm. So far, the method has only been applied to 3D electron gas, which
  is a paradigmatic system, but a rather simple one. Here, we systematically extend the
  method to 2D including realistic systems such as mono and bilayer graphene. We focus on
  one of the most important vertex function effects involving the exchange of two
  particles in the final state. We demonstrate that it should be evaluated with the proper
  screening and discuss its influence on the quasiparticle properties.
  %-----------------------------------------------------------------------------------------
\end{abstract}
\maketitle
%=======================================================================================
%                                Sec. I. INTRODUCTION          ////////////////////////
%=======================================================================================
\section{Introduction}
\begin{figure}[b] 
  \centering \includegraphics[width=\columnwidth]{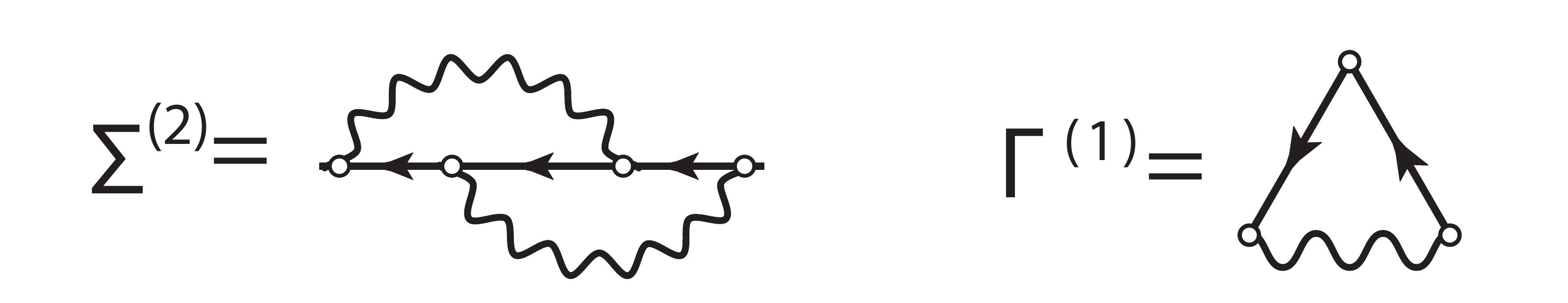}
\caption{\label{fig:SGM2} A single second-order self-energy diagram and the associated
  first-order vertex function in terms of the electron propagators (arrows) and the
  screened interactions (wavy lines).}
\end{figure}

Numerous correlated electron calculations follow a canonical scheme formulated by
Hedin~\cite{hedin_new_1965} in terms of dressed propagators. It is now well established
that the lowest-order self-energy (SE) term, the so-called $GW$ approximation is the major
source of electronic correlations. Much less is known about the next perturbative orders:
there is no single standard way of evaluating them despite the fact that there is a single
second-order self-energy diagram (Fig.~\ref{fig:SGM2}). There are multiple reasons for
this.  On one side, at the advent of many-body perturbation theory (MBPT) the
computational power was insufficient to perform these demanding calculations, and one was
forced to use some drastic simplifications. On the other side, there are several
conceptual problems with the organization of many-body perturbation theory (MBPT) for
interacting electrons. For instance, it is known that higher-order diagrammatic
approximations for the electron self-energy in terms of the screened Coulomb interaction
$W$ leads to poles in the ``wrong'' part of the complex plane giving rise to
\emph{negative spectral densities}. This observation has been made long time ago by
Minnhagen~\cite{minnhagen_vertex_1974,minnhagen_aspects_1975}, and in our recent works we
provided a general solution to this
problem~\cite{stefanucci_diagrammatic_2014,uimonen_diagrammatic_2015} yielding positive
definite (PSD) spectral functions. The idea was to write the self-energy in the Fermi
Golden rule form well known from the scattering theory.

One interesting conclusion of our theory is that the second-order SE describes three
distinct scattering processes that take place in a many-body
system~\cite{pavlyukh_vertex_2016}: (I) A correction to the first-order scattering,
involving the same final states as in $GW$. This effect was numerically studied in
Ref.~\cite{stefanucci_diagrammatic_2014}, and has been shown~\cite{pavlyukh_vertex_2016}
to counteract the smearing out of spectral features in self-consistent
calculations~\cite{holm_fully_1998}. (II) Excitation of two plasmons ($pl$), or two
particle-hole pairs ($p$-$h$), or a mixture of them in the final state. Especially the
generation of two plasmons is a prominent effect spectroscopically manifested as a second
satellite in the photoemission spectrum~\cite{riley_crossover_2018}. This effect can be
obtained from the cumulant expansion~\cite{holm_self-consistent_1997,
  guzzo_multiple_2014}, which, however, only works at the band bottom $k=0$, or from the
Langreth model~\cite{langreth_singularities_1970,pavlyukh_pade_2017}. (III) A first-order
scattering involving the exchange of the two final state particles. This latter scattering
process is the focus of the present work.

Some manifestations of the mechanism (III) have already been studied, albeit without
realizing its deep connection with the full $\Sigma^{(2)}$. First of all, for the two
\emph{bare} interaction lines we get the so-called \emph{second-order exchange}, which has
been shown to play an important role in correlated electronic calculations for molecular
systems as an ingredient of the second-Born approximation
(2BA)~\cite{balzer_time-dependent_2010,perfetto_first-principles_2015,schuler_spectral_2018}.
Second, it yields a very important total energy correction for the homogeneous electron
gas~\cite{ziesche_self-energy_2007}. Third, the mechanism \emph{with screening} has been
considered in the calculations of quasiparticle life-times. Reizer and Wilkins predicted
that this diagram yields a 50\% reduction of the scattering rate in 2D electron gas
calling it ``a nongolden-rule'' contribution, whereas Qian and
Vignale~\cite{qian_lifetime_2005} correctly pointed out that it is ``still described by
the Fermi golden rule, provided one recognizes that the initial and final states are
Slater determinants'', and that the coefficient is different.  Fourth, the mechanism is
relevant for the scattering theory~\cite{almbladh_photoemission_2006}. With bare Coulomb
interactions it represents the so-called double photoemission (DPE) process, and if the
interaction is screened\,---\,the plasmon assisted
DPE~\cite{pavlyukh_single-_2015,schuler_electron_2016}.  Finally, the considered mechanism
has some features in common with the second-order screened exchange (SOSEX)
approximation~\cite{gruneis_making_2009,ren_beyond_2015}. However, there are also
important differences in the constituent screened Coulomb interaction that will be
explained below.

As can be seen from this list, the mechanism is an indispensable part of various physical
processes. However, it has not been sufficiently emphasized that all of them can be
derived from a single $\Sigma^{(2)}$ diagram. Moreover, there are no systematic studies of
its impact on the quasiparticle properties other than the life-times. These gaps are
filled in here. Our theoretical derivations are illustrated by calculations for four
prominent systems: the homogeneous electron gas in two and three dimensions and the mono-
and bilayer graphene. While the former two are very well studied model
systems~\cite{lundqvist_single-particle_1968, santoro_electron_1989}, graphene is a real
material, and while the $GW$ calculations for it exists~\cite{hwang_quasiparticle_2008,
  polini_plasmons_2008, sensarma_quasiparticles_2011}, MBPT has mostly been used in the
renormalization group sense~\cite{kotov_electron-electron_2012}. Little is known about the
frequency dependence of higher-order self-energies.

Our approach consists of analytical and numerical parts. For the quasiparticle ($qp$)
electron Green's function ($G_0$) and the screened interaction ($W_0$) in the random phase
approximation (RPA), the frequency integration of a selected set of the electron
self-energy ($\Sigma[G_0,W_0]$) diagrams is performed in closed form using our symbolic
algorithm implemented in \textsc{mathematica} computer algebra system. The remaining
momentum integrals are performed numerically in line with our previous studies using the
Monte Carlo
approach~\cite{pavlyukh_time_2013,stefanucci_diagrammatic_2014,uimonen_diagrammatic_2015,pavlyukh_vertex_2016}
showing excellent accuracy and scalability. First, we evaluate the scattering rate
function
\begin{align}
  \Gamma(k,\omega)&=\ii\left[\Sigma_c^>(k,\omega)-\Sigma_c^<(k,\omega)\right],
  \label{eq:rate}
\end{align}
and then the retarded self-energy via the Hilbert transform (Appendix~\ref{sec:app:hilbert})
\begin{align}
  \SgR(k,\omega)&=\Sigma_x(k)+
  \int\frac{d\omega'}{2\pi}\frac{\Gamma\mleft(k,\omega'\mright)}{\omega-\omega'+\ii \eta},
  \label{eq:re:sgm:R}
\end{align}
where $\Sigma_x(k)$ is the frequency-independent exchange self-energy, and the meaning of
greater $>$ and lesser $<$ components of the correlated self-energy $\Sigma_c$ is
explained in the next section. Via the Dyson equation (Appendix~\ref{app:dyson}), the
retarded self-energy determines correlated electronic structure.

Our work is structured as follows: we review our PSD approach in Sec.~\ref{sec:psd} and
illustrate it with a concrete set of diagrams in Sec.~\ref{sec:diag}. Next we discuss the
building blocks of our diagrammatic perturbation theory and provide reference $G_0W_0$
calculations for the four systems in Sec.~\ref{sec:systems}.  Efficient evaluation of
screening is an important ingredient. In Sec.~\ref{sec:SGM:aA} we present our main
numerical results: spectral features in $\Sigma^{(2)}$, cancellations between the first
and the second order self-energies in the asymptotic regime, quasiparticle properties such
as quasiparticle peak strengths, effective masses, velocities, and life-times. We
finally present our conclusions and outlooks in Sec.~\ref{sec:conclusions}.

\section{Summary of the PSD approach\label{sec:psd}}
Besides numerical difficulties, the major reason on why the MBPT calculations for the
electron gas have not been systematically performed at higher orders is the fact that
resulting expansions do not generate positive definite (PSD) spectral functions at all
frequency and momentum values. How to deal with this obstacle is discussed in details in
Refs.~\cite{stefanucci_diagrammatic_2014,uimonen_diagrammatic_2015}.

Even though this is an equilibrium problem, our method can most easily be formulated by
using the nonequilibrium Green's function (NEGF)
formalism~\cite{stefanucci_nonequilibrium_2013}. The main distinction is that field
operators ($\hpsi(\bx,z)$ and $\hpsi^\dagger(\bx,z)$ for electrons) evolve on the
time-loop contour $z\in\mC$ with one forward chronologically ordered ($\mC-$) branch and
one ($\mC^+$) branch with anti-chronological time-ordering,
$\mC=\mC^-\cup\mC^+$. Correspondingly, the two times Green's functions generalize to
$G(\bx_1z_1,\bx_2z_2)$ or $G^{\alpha\beta}(\bx_1t_1,\bx_2t_2)$, where $t_1$ and $t_2$ are
the projections of $z_{1,2}$ on the real time-axis, and $\alpha,\beta=+/-$ indicate to
which branches of the Keldysh contour they belong. In the following, we will explicitly
deal with the \emph{lesser} self-energy $\Sigma^<\equiv\Sigma^{-+}$, which describes
scattering processes on the subspace of states below the Fermi level, i.\,e., holes. The
\emph{greater} component ($\Sigma^>\equiv\Sigma^{+-}$) can be treated analogously.

\begin{figure}[t] 
  \centering
\includegraphics[width=\columnwidth]{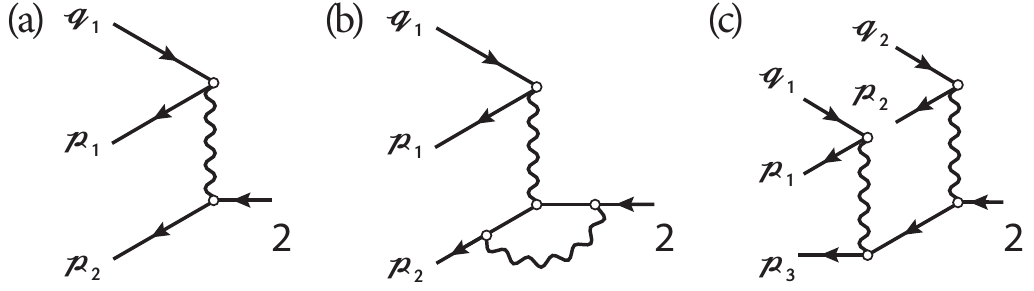}
\caption{\label{fig:diag1} Half-diagrams for $D(2)$, the constituent of the
  $\Sigma^<(1,2)$ SE. All vertices are on the $\mC^{+}$ branch. Wavy lines stand for the
  screened interaction $W$.}
\end{figure}

The PSD property concerns the fact that the rate operator~\eqref{eq:rate} must be positive
for all momentum $k$ and frequency $\omega$ values. $\Sigma_c$ with this property is
diagrammatically constructed starting from any given set of diagrams as follows.

On the first step pluses and minuses are assigned to the diagram vertices in all possible
combinations. They carry information about the contour times. The resulting decorated
diagrams are called \emph{partitions}. Since we have shown that at zero temperature no
isolated $+$ or $-$ islands can exist~\cite{stefanucci_diagrammatic_2014}, the ``cutting''
procedure splits the diagrams for $\Sigma_c^<$ into halves that have their vertices
exclusively either on the ($-$) or on the ($+$) branch (viz. Fig.~\ref{fig:diag1}). They
are the building blocks of the PSD construction. Subsequently, the half-diagrams are
combined in such a way that a sum of complete squares is formed. This guarantees the
positivity of the resulting set of diagrams. On the language of scattering theory, the
half-diagrams have a meaning of the $S$-matrices describing various particle or hole
scattering processes in a many-body system. The resulting PSD self-energies have then the
Fermi Golden rule form, which always leads to positive scattering rates. Topologically
distinct $S$-matrices will be denoted as $D$ diagrams. Diagrams that can be obtained by
the cutting procedure applied to $\Sigma_c$ of the first- and second-order in $W_0$ are
depicted in Fig.~\ref{fig:diag1}.

They are interpreted according to the standard diagrammatic rules.  Consider for instance
the half-diagrams with all the time-arguments on the $+$ branch (such as depicted in
Fig.~\ref{fig:diag1}).  In addition to the initial one-hole ($1h$) state (with the
coordinate $2$, where the composite position-spin and time variables are abbreviated as
$i\equiv(\bx_i,t_i)$), the final state is denoted by the two strings of numbers
$\cpu=(\cp_1,\ldots \cp_N, \cp_{N+1})$ and $\cqu=(\cq_1,\ldots \cq_N)$ that specify composite
coordinates of the outgoing $N+1$ holes and $N$ particles, respectively. We further
associate a single time-argument $\tau$ with $(\cpu,\cqu)$. $\tau$ is the latest time on the
forward and the earliest time on the backward contour branches. With these notations,
$D^{(a)}$ reads
\begin{subequations}
\begin{align}
  D^{(a)}_{\cp_1,\cp_2,\cq_1}(2)&=-(-1)^{1}\int\!\dd(4)\,W_0^{++}(4,2)g^<(\cp_2\tau,2)\nn\\
  &\quad\times g^<(\cp_1\tau,4)g^>(4,\cq_1\tau),\label{eq:D2}
\end{align}
and its complex conjugate is given by
\begin{align}
  \big[D^{(a)}_{\cp_1,\cp_2,\cq_1}(1)\big]^*&=+\int\!\dd(3)\,W_0^{--}(1,3)g^<(1,\cp_2\tau)\nn\\
  &\quad\times g^<(3,\cp_1\tau)g^>(\cq_1\tau,3),\label{eq:D1}
\end{align}
\label{eq:D}
\end{subequations}
where the extra minus sign $(-1)^{1}$ in Eq.~\eqref{eq:D2} is due to the fact that for
each time-integration associated with a vertex on $\mC^+$
\begin{align}
 \int_{\mC^+} \dd z_i\ldots&=-\int_{-\infty}^{\infty} \dd t_i\ldots.
\end{align}
Eqs.~\eqref{eq:D} are expressed in terms of the bare electron propagators
$g(\bx_1,t_1;\bx_2,t_2)$ and the RPA screened interaction 
\begin{align}
  W_0(1,2)&=\int \dd(3) v(1,3)\,\varepsilon_0^{-1}(3,2),
\end{align}
where $\varepsilon_0$ is the RPA dielectric function defined in terms of the polarization
bubble $\mP_0$ and the bare Coulomb interaction $v$,
\begin{align}
  \varepsilon_0(1,2)&=\delta(1,2)-\int\dd(3)\, v(1,3)\mP_0(3,2),\\
  \mP_0(1,2)&=-\ii g(1,2)g(2,1).
\end{align}
In Eqs.~\eqref{eq:D}, $W_0^{--}\equiv \WnT$ and $W_0^{++}\equiv \WnAT$ stand for the
\emph{time-ordered} and \emph{anti-time-ordered} interactions, respectively. We refer to
App.~\ref{sec:app:prop} for the detailed definitions and Sec.~\ref{sec:systems} for
explicit forms of the dielectric function for the four studied systems. $D^{(b)}$ and
$D^{(c)}$ are defined analogously. Our next goal is to describe self-energies that are
obtained by ``gluing'' the $D$-diagrams. This is complementary to our earlier
works~\cite{stefanucci_diagrammatic_2014,uimonen_diagrammatic_2015}, where the
half-diagrams were derived by the ``cutting'' rules.
%=======================================================================================
%                           Sec. III. Physical interpretation   ////////////////////////
%=======================================================================================
\section{Self-energy approximations: physical meaning of diagrams\label{sec:diag}}
\begin{figure}[] 
  \centering
\includegraphics[width=\columnwidth]{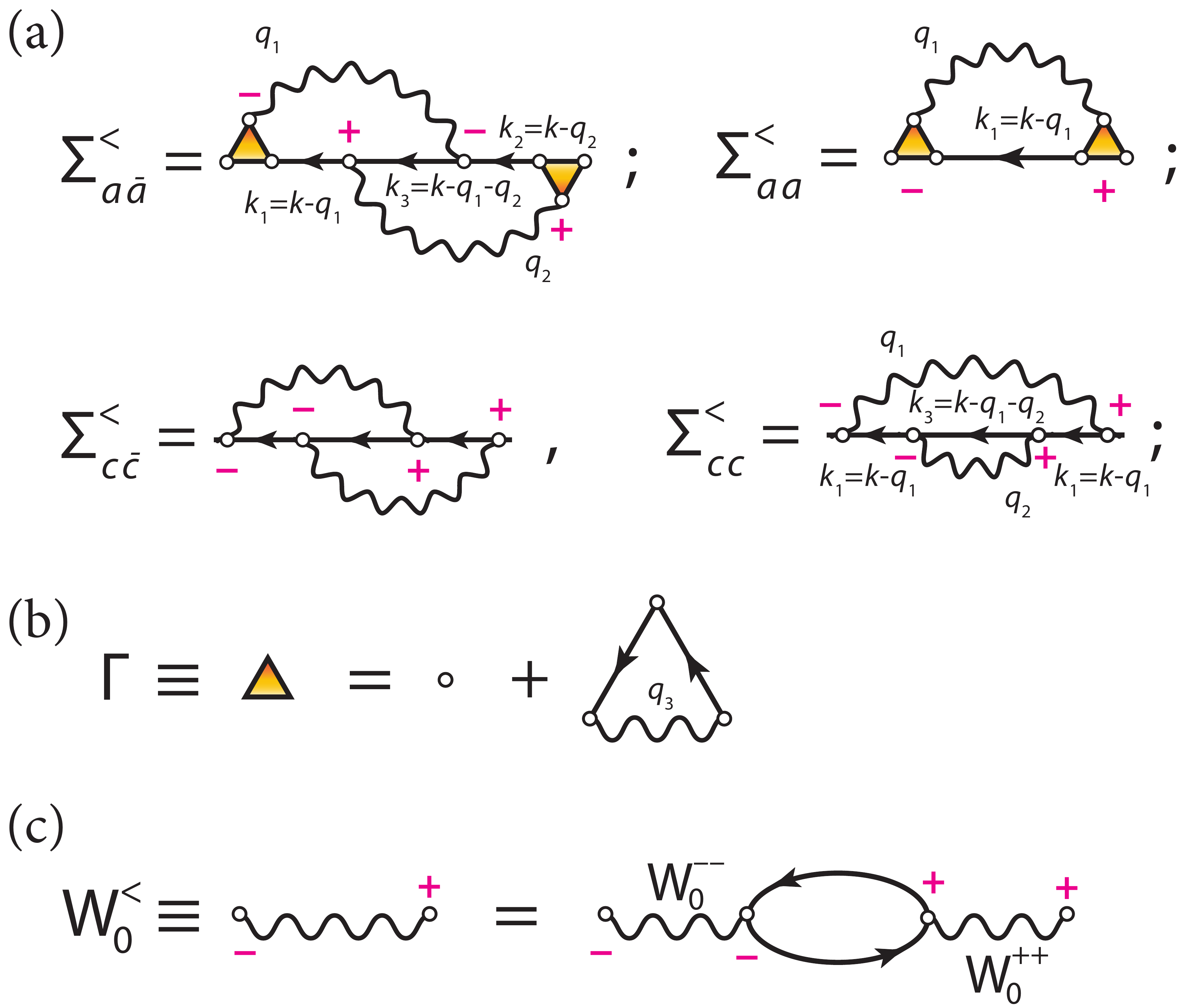}
\caption{\label{fig:diag2} (a) The contributions to $\Sigma_\text{PSD}^<$ given by
  Eq.~\eqref{eq:sgm_abc} and arising from the first- and second-order (in screened
  interaction) self-energy by virtue of the PSD procedure (Sec.~\ref{sec:psd}). Three
  subsets of these diagrams that also fulfill the PSD property are given by
  Eqs.~\eqref{eq:subs:sgm:psd}. (b) First order vertex function $\mathsf{\Gamma}$.  (c)
  Random phase approximation for the lesser component of the screened Coulomb
  interaction.}
\end{figure}

It is straightforward to see that by `gluing'' three half-diagrams $D^\text{(i)}(2)$
($\text{i}=a,b,c$, Fig.~\ref{fig:diag1}) with their complex conjugates
$[D^\text{(i)}(1)]^*$ with or without permutations of internal coordinates, one obtains
the four classes shown in Fig.~\ref{fig:diag2}(a). They are grouped into three terms
covering three distinct physical mechanisms
\begin{equation}
  \Sigma_\text{PSD}^<(1,2)=\Sigma^<_{aa}(1,2)+[\Sigma^<_{cc}(1,2)+\Sigma^<_{c\bar{c}}(1,2)]
  +\Sigma^<_{a\bar{a}}(1,2),
  \label{eq:sgm_abc}
\end{equation}
In Ref.~\cite{stefanucci_diagrammatic_2014} we have also shown that this is \emph{the
  minimal set} of diagrams covering all the first- and second-order self-energies and
possessing the PSD property. Let us discuss the involved physical mechanisms and derive
the working formulas.

%=============================================================
%            Sec. III.A. 3 basic mechanisms
%=============================================================
\subsection{$\Sigma_{aa}^<$}
$\Sigma_{aa}^<$ without vertex corrections is nothing else as the first-order ($GW$)
self-energy. It results from the gluing the simplest half-diagram $D^{(a)}$
[Fig.~\ref{fig:diag1}(a)] with itself without permuting the two hole lines ($\cp_1$ and
$\cp_2$):
\begin{align}
  \Sigma_{GW}^<(1,2)&=\ii \sum_{\cp_1,\cp_2,\cq_1}\big[D_{\cp_1,\cp_2,\cq_1}^{(a)}(1)\big]^*
  D^{(a)}_{\cp_1,\cp_2,\cq_1}(2)\nn\\
  &=\ii g^<(1,2)W_0^<(1,2).\label{eq:GW}
\end{align}
In order to establish the second equality, we use the explicit form of the
half-diagrams~\eqref{eq:D}, recall that
$\ii g^{<}(1,2)=\sum_{\cp_2}g^<(1,\cp_2\tau)g^<(\cp_2\tau,2)$, $\ii g^<(3,4)=\sum_{\cp_1}
g^<(3,\cp_1\tau)g^<(\cp_1\tau,4)$, and $-\ii g^>(4,3)=\sum_{\cq_1}g^>(4,\cq_1\tau)g^>(\cq_1\tau,3)$,
and that the lesser screened interaction can be written in the form
\begin{align*}
  W_0^<(1,2)&=-\!\iint \!\!\dd(3,4)\, W_0^{--}(1,3)\mP^{<}_0(3,4)W_0^{++}(4,2),
\end{align*}
with $\mP^{<}_0(3,4)=-\ii g^<(3,4)g^>(4,3)$ as shown in Fig.~\ref{fig:diag2}(c).

Now we use the diagrams in momentum and frequency representation as indicated in
Fig.~\ref{fig:diag2}, namely
\begin{align}
  \vec{k}_{1}&=\vec{k}-\vec{q}_{1}, &\vec{k}_{2}&=\vec{k}-\vec{q}_{2},&\vec{k}_{3}&=\vec{k}-\vec{q}_{1}-\vec{q}_{2},
  \label{eq:momentum:cons}
\end{align}
in order to derive a standard result for the $GW$ self-energy:
\begin{align}
\Sigma_{GW}^<(k,\w)&=\ii\int \dd \Omega_1 \int \frac{\dd\nu_1}{2\pi}\,
g^<(k_{1},\w-\nu_1)\nn\\
&\qquad\times W_0^<(q_{1},\nu_1),\label{sgm_gw}
\end{align}
where $\int\dd \Omega_1\equiv\int \frac{\dd^d\vec{q}_{1}}{(2\pi)^d}$ denotes an integral over
a $d$-dimensional momentum space, $\w$ is the external frequency and $\vec{k}$ is the
momentum. For graphene systems, the integration additionally contains a sum over the bands
and a respective scattering matrix element. We will generally use $\w_i$ and $\vec{k}_{i}$ for
the energy and momentum of fermionic lines, and $\nu_i$ and $\vec{q}_{i}$ for the interaction
lines.

Introducing the spectral function of the screened interaction $C(q,\nu)$ and using
explicit formulas for the bare propagators in Appendix~\ref{sec:app:prop} and in
particular
\begin{align}
  g^<(k_{},\w)&=2\pi\ii \nF(k_{}) \dlf(\w-\epsilon(k_{})),\\
  W_0^{<}(q_{},\nu)&=- 2\pi \ii \thf(-\nu) C(q_{},-\nu),\label{eq:W:lss}
\end{align}
where $\nF(k_{})\equiv\nF(\epsilon(k_{}))$ is the fermion occupation number, we obtain
\begin{align}
  \Sigma_{GW}^{<}(k,\w)&=2\ii\pi\int \dd \Omega_1 \int_0^\infty \dd\nu_1\,
  \nF(k_{1})C(q_{1},\nu_1)\nn\\
  &\qquad\times \dlf(\w+\nu_1-\epsilon_1),
  \label{eq:sgm:aa}
\end{align}
with $\epsilon_i\equiv\epsilon(\vec{k}_{i})$.  As long as the spectral function of neutral
excitations is positive, $C(q_{},\nu)>0$ (which is indeed the case because we use RPA for
$W_0$ here~\cite{uimonen_diagrammatic_2015}), the rate operator
[$-\ii \Sigma_{GW}^<(k,\w)$] is positive too, as evident from Eq.~\eqref{eq:sgm:aa}, and
the nature of the final scattering state is revealed: it consists of a hole with energy
$\epsilon_1-\nu_1$ and a neutral excitation such as $p$-$h$ pair or a plasmon with
momentum $q_{1}$ and energy $\nu_1$.

Eq.~\eqref{eq:GW} can be extended by adding internal interaction lines to $D^{(a)}$ and
maintaining the external indices and the way how the constituent half-diagrams are
glued. The $b$-half-diagram depicted in Fig.~\ref{fig:diag1}(b) represents the simplest
possibility
\begin{align}
\Sigma_{aa}^<(1,2)&=\ii \sum_{\cp_1,\cp_2,\cq_1}\big[D^{(a)}+D^{(b)}\big]_{\cp_1,\cp_2,\cq_1}^*(1)\nn\\
&\quad\quad\quad\times\big[D^{(a)}+D^{(b)}\big]_{\cp_1,\cp_2,\cq_1}(2).
\label{eq:sgm_aa}
\end{align}
$D^{(b)}$ has one extra interaction line and therefore by gluing it with $D^{(a)}$ leads
to two equivalent terms of the second order in $W_0$, and by gluing $D^{(b)}$ with itself
to a term of third order. They can conveniently be represented by introducing the vertex
function $\mathsf\Gamma$ depicted as yellow triangle in Fig.~\ref{fig:diag2}\,(a,b) and
familiar from the Hedin's functional
equations~\cite{hedin_new_1965,strinati_application_1988}. If one starts from higher-order
diagrams, the diagrammatic expansion of $\mathsf\Gamma$ becomes more complicated and
starts to differ from the standard vertex function\footnote{Since our theory maintains the
  Fermi Golden rule form, $\mathsf\Gamma$ enters symmetrically unlike in the Hedin's
  theory. Not surprisingly, such object was introduced for the first time in the context
  of photoemission by Almbladh~\cite{almbladh_photoemission_2006}.}. As in the case of
$\Sigma_{GW}$, the electronic and the interaction lines connecting the $+$ and $-$ islands
are given by the lesser propagators $g^<(k_{1},\w-\nu_1)$ and $W^<(q_{1},\nu_1)$. In view
of the energy conservation, only these two propagators depend on $\nu_1$, and the
frequency integration can likewise be performed.  It is clear that the same functional
form proportional to $\nF(\w+\nu_1)\delta(\w-\epsilon_1+\nu_1)$ is obtained. Therefore, we
conclude that $D^{(b)}$ renormalizes the $GW$ expression, but does not lead to new
spectral features. Eq.~\eqref{eq:sgm_aa} is a complete square, therefore
$\Sigma_{aa}^<(1,2)$ is PSD. It was numerically evaluated in our earlier
work~\cite{stefanucci_diagrammatic_2014}.

\subsection{$\Sigma_{c\bar{c}}^<$ and $\Sigma_{cc}^<$}
The same analysis can be applied to other diagrams. $\Sigma_{c\bar{c}}^<$ and
$\Sigma_{cc}^<$ feature the $--++$ partition (in this notation the vertices are traversed
along the fermionic lines from 1 to 2 in the order opposite to arrows) and contain two
diagrams from gluing the half-diagrams of the $c$-type [Fig.~\ref{fig:diag1}(c)]
$[D^{(c)}(1)]^*$ and $D^{(c)}(2)$ with and without permutation of the dangling fermionic
lines, respectively:
\begin{multline}
\Sigma_{cc}^<(1,2)+\Sigma_{c\bar{c}}^<(1,2)=\ii \sum_{\cpu,\cqu}
      \big[D^{(c)}_{\cp_1,\cp_2,\cp_3,\cq_1,\cq_2}(1)\\
        +D^{(c)}_{\cp_2,\cp_1,\cp_3,\cq_2,\cq_1}(1)\big]^*D^{(c)}_{\cp_1,\cp_2,\cp_3,\cq_1,\cq_2}(2).
      \label{eq:sgm_cc}
\end{multline}
Explicit derivation of this expression in momentum-energy representation goes beyond the
scope of this work. However, some physical insight can be gained by using the plasmon-pole
approximation $C(q,\nu)=C(q)\delta(\nu-\Omega(q))$ for the screened
interaction~\eqref{eq:W:lss}. As can be seen from the diagrammatic representation
(Fig.~\ref{fig:diag2}) of self-energy~\eqref{eq:sgm_cc}, there are 3 lesser propagators
connecting the $+$ and $-$ islands, which in the energy-momentum representation read:
$g^<(k_{3},\w-\nu_1-\nu_2)W^<(q_{1},\nu_1)W^<(q_{2},\nu_2)$. In view of the energy
conservation these are the only propagators that depend on the frequencies
$\nu_{1,2}$. Therefore, the integrals can be explicitly performed. A scattering process
accompanied by the generation of two plasmons can be inferred from the resulting frequency
dependence proportional to
$\nF(\w+\Omega(q_1)+\Omega(q_2))\delta(\w-\epsilon_3+\Omega(q_1)+\Omega(q_2))$, and is PSD per
construction.

\begin{figure}[t] 
  \centering
  \includegraphics[width=\columnwidth]{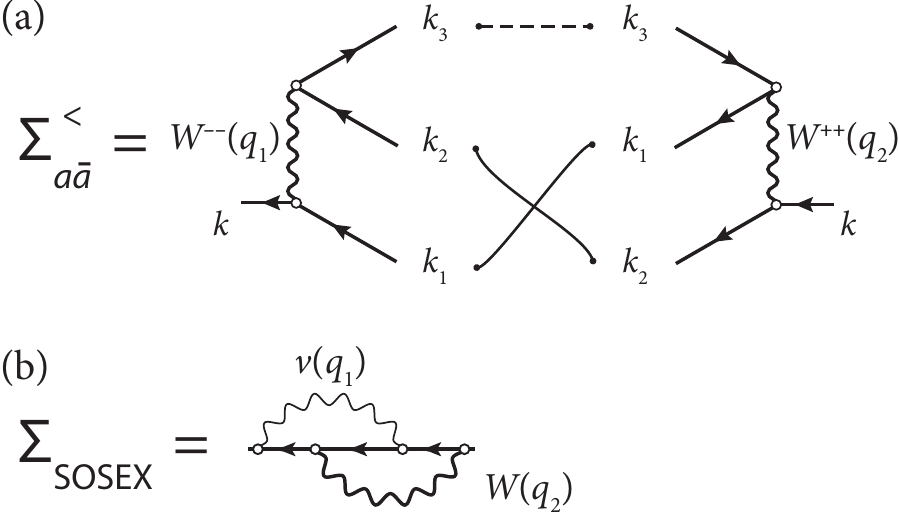}
\caption{\label{fig:SGMaA} (a) Gluing two $D^{(a)}$ half-diagrams with one permutation of
  the two hole lines with momenta $\vec{k}_{1}=\vec{k}-\vec{q}_{1}$ and
  $\vec{k}_{2}=\vec{k}-\vec{q}_{2}$ and a single hole line
  $\vec{k}_{3}=\vec{k}-\vec{q}_{1}-\vec{q}_{2}$ (dashed) yields
  $\Sigma_{a\bar{a}}^<(k,\w)$. The unpermuted configuration is not included, therefore the
  diagram may lead to a negative electron spectral function. Wavy lines denote screened
  Coulomb interactions. (b) $\Sigma_\text{SOSEX}^<$ contains different interaction lines
  and is thus distinct from $\Sigma_{a\bar{a}}^<$.}
\end{figure}
\subsection{$\Sigma_{a\bar{a}}^<$}
Finally we consider a rather complicated $\Sigma_{a\bar{a}}^<$ resulting from the $-+-+$
partition, Fig.~\ref{fig:diag2}\,(a):
\begin{align}
  \Sigma_{a\bar{a}}^<(1,2)&=-\ii \sum_{\cp_1,\cp_2,\cq_1}\!\big[D^{(a)}+D^{(b)}\big]_{\cp_2,\cp_1,\cq_1}^*(1)\nn\\
  &\qquad\qquad\times\big[D^{(a)}+D^{(b)}\big]_{\cp_1,\cp_2,\cq_1}(2).
\label{eq:sgm_aA}
\end{align}
It has a form very similar to Eq.~\eqref{eq:sgm_aa}, except the hole indices $\cp_1$ and
$\cp_2$ are permuted (Fig.~\ref{fig:SGMaA}) leading to the change of sign. The sign of a
permutation can be conveniently determined from the number of crossing of fermionic lines
connecting the half-diagrams~\cite{van_leeuwen_wick_2012}. Neglecting the $D^{(b)}$
diagrams, which only produces a correction to the scattering of a hole state into a
2-holes-1-particle state, and using the explicit form for $D^{(a)}$, Eqs.~\eqref{eq:D},
the self-energy in coordinate representation reads
\begin{align}
  \Sigma_{a\bar{a}}^<(1,2)=-\iint \dd(3,4)& \, W^{--}(1,3) g^<(1,4)g^>(4,3)\nn\\
  &\quad\times  g^<(3,2) W^{++}(4,2).
\end{align}
Thus, there are 2 lesser and 1 greater propagators connecting the $+$ and the $-$
islands. In the momentum-frequency representation they are
$g^>(k_{3},\w-\nu_1-\nu_2)g^<(k_{1},\w-\nu_1)g^<(k_{2},\w-\nu_2)$.  They
contain $\delta$-functions, therefore, the integrals over the internal frequencies
$\nu_{1,2}$ are simple.  Collecting screened interaction dependent on these frequencies,
$W^{--}(q_{1},\nu_1) W^{++}(q_{2},\nu_2)$, and using the explicit form
\begin{align}
  g^>(k_{},\w)&=-2\pi \ii \nbF(k_{})\delta(\w-\epsilon(k_{})),
\end{align}
with $\nbF(k_{})=1-\nF(k_{})$, we can write the self-energy explicitly
\begin{align}
  \Sigma_{a\bar{a}}^<(k,\w)&=2\ii\pi\!\iint\!\dd(\Omega_{1,2})\,
  \nF(\epsilon_1)\nF(\epsilon_2)\nbF(\epsilon_3) \nn\\ &\qquad\times
  W^{--}(q_{1},\w-\epsilon_1)
  W^{++}(q_{2},\w-\epsilon_2)\nn\\ &\qquad\times\delta(\w-\epsilon_1-\epsilon_2+\epsilon_3)
  \label{eq:sgm:aA}\\
  &=-2\ii\pi\!\iint\!\dd(\Omega_{1,2})\,
  \nF(\epsilon_1)\nF(\epsilon_2)\nbF(\epsilon_3)   \nn\\
  &\qquad\times \re\left[\WT(q_{1},\w-\epsilon_1)
    \left(\WT(q_{2},\w-\epsilon_2)\right)^*\right]\nn\\
  &\qquad\times\delta(\w-\epsilon_1-\epsilon_2+\epsilon_3).
\end{align}

%\subsection{Physical meaning of $\Sigma_{a\bar{a}}$}
Because of the permutation of $\cp_1$ and $\cp_2$ indices, Eq.~\eqref{eq:sgm_aA} forms a
complete square only in combination with the unpermuted configuration,
Eq.~\eqref{eq:sgm_aa}, and $-\ii \Sigma_{a\bar{a}}^<$ is not PSD on its own. At least for
bare interactions, this can immediately be seen from the equation above. In this case the
second-order exchange self-energy $\Sigma_{2x}$ is obtained.

Because $\Sigma_{2x}$ is a limit of $\Sigma_{a\bar{a}}$, one might call the latter as a
second-order screened exchange (SOSEX). However, this is not the common definition and
therefore we will use $\Sigma_{a\bar{a}}$ to contrast it with $\Sigma_\text{SOSEX}$. So,
what is the difference between the two? $\Sigma_\text{SOSEX}$ has been derived by
Freeman~\cite{freeman_coupled-cluster_1977} and applied to the computation of total
energies by Gr{\"u}neis \emph{et al.}~\cite{gruneis_making_2009} and spectral properties
by Ren \emph{et al.}~\cite{ren_beyond_2015}. The starting point is the screened
interaction in the RPA form
\begin{align}
  W(1,2)&=v(1,2)+\int\dd(3,4)v(1,3)\mP_0(3,4)W(4,2).
\end{align}
$\Sigma_\text{SOSEX}$ is obtained by inserting the second term in the $GW$ self-energy and
interchanging the two electron propagators. As can be seen from Fig.~\ref{fig:SGMaA}(b),
one constituent interaction is bare, whereas another one is screened. This is to be
contrasted with $\Sigma_{a\bar{a}}$, where both lines are screened. Notice, there is no
double counting because they belong to different branches of the Keldysh contour.

%=======================================================================================
%                                Sec. IV. SYSTEMS               ////////////////////////
%=======================================================================================
\section{Systems and reference results\label{sec:systems}}
\begin{figure}[t] 
  \centering
    \includegraphics[width=\columnwidth]{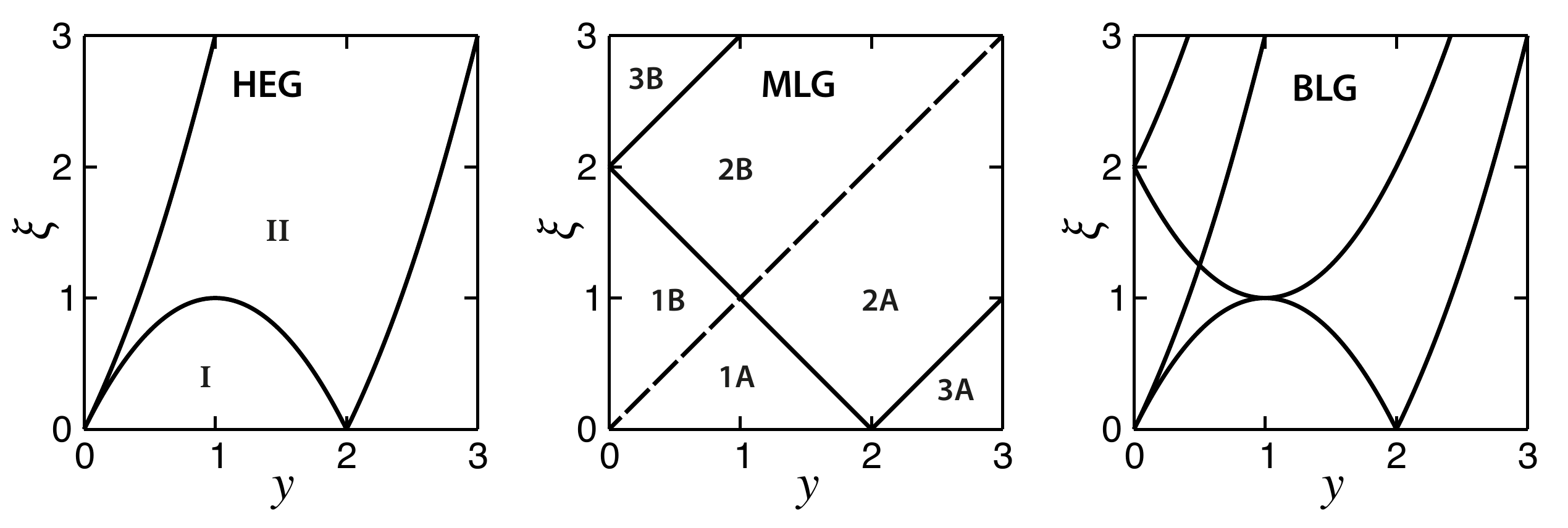}
\caption{\label{fig:chi:dom} Domains in the momentum-energy plane in the definitions of
  the dielectric functions of HEG, MLG, and BLG. Here $y=q/\kF$ and $\xi=\w/\eF$.}
\end{figure}
\begin{figure}[t] 
  \centering
    \includegraphics[width=\columnwidth]{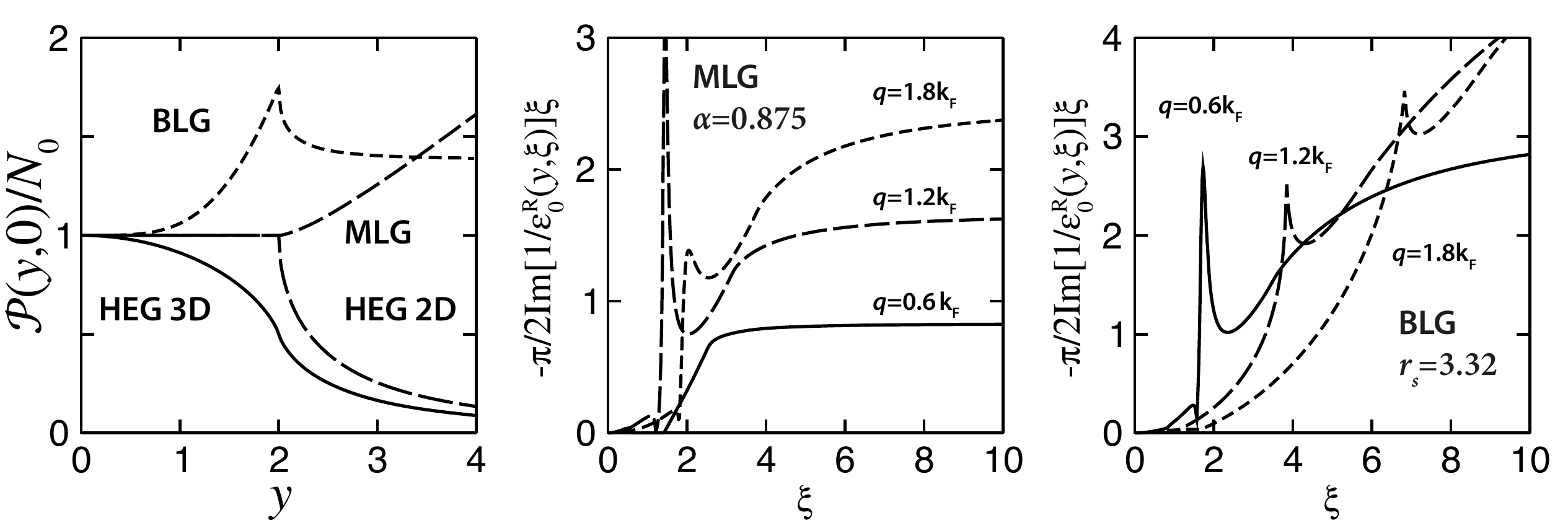}
\caption{\label{fig:chi:k:fsum} Static irreducible polarization $\mP(q,0)$ for the four
  studied systems normalized at the density of states at the Fermi level $N_0$ (a). $N_0$
  is given by $\rho_\sigma(\eF)$ for HEG and by $\rho_{\sigma,s}(\eF)$ for the two
  graphene systems, $y=q/\kF$ and $\xi=\w/\eF$. Integrand of the $f$-sum rule for MLG (b)
  and BLG (c) demonstrating the divergence of the $f$-sum in these systems. }
\end{figure}

In this section we present in a uniform way the four studied systems. We focus on the
dielectric function in the momentum-frequency plane, Fig.~\ref{fig:chi:dom}. It is closely
connected to the irreducible polarization $\mP(q,\w)$ and to the density-density response
$\chi(q,\w)$,
\begin{align}
  \chi(q,\w)&=\mP(q,\w)+\mP(q,\w) v(q) \chi(q,\w).
  \label{dyson}
\end{align}  
They determine the microscopic dielectric function and its inverse, respectively,
\begin{align}
  \varepsilon(q,\w)&=1-g v(q) \mP(q,\w),\\
  \varepsilon(q,\w)^{-1}&=1+g v(q) \chi(q,\w).\label{eq:eps:def}
\end{align}
Here $g$ is the degeneracy factor. For the homogeneous electron gas there is only spin
degeneracy, $g=g_s=2$, whereas for the mono- and bilayer graphene the valley degeneracy
additionally appears $g=g_sg_v$. For these systems $g_v=2$, but it can take larger values
for other systems~\cite{das_sarma_electronic_2011,basov_polaritons_2016}. The density of
states at the Fermi energy $N_0$ is a natural unit to measure $\mP$ and $\chi$, because
the static polarization $\mP(q,0)$ for small values of $q$ is exactly given by this
quantity, Fig.~\ref{fig:chi:k:fsum}(a).

\begin{figure}[] 
  \centering
    \includegraphics[width=\columnwidth]{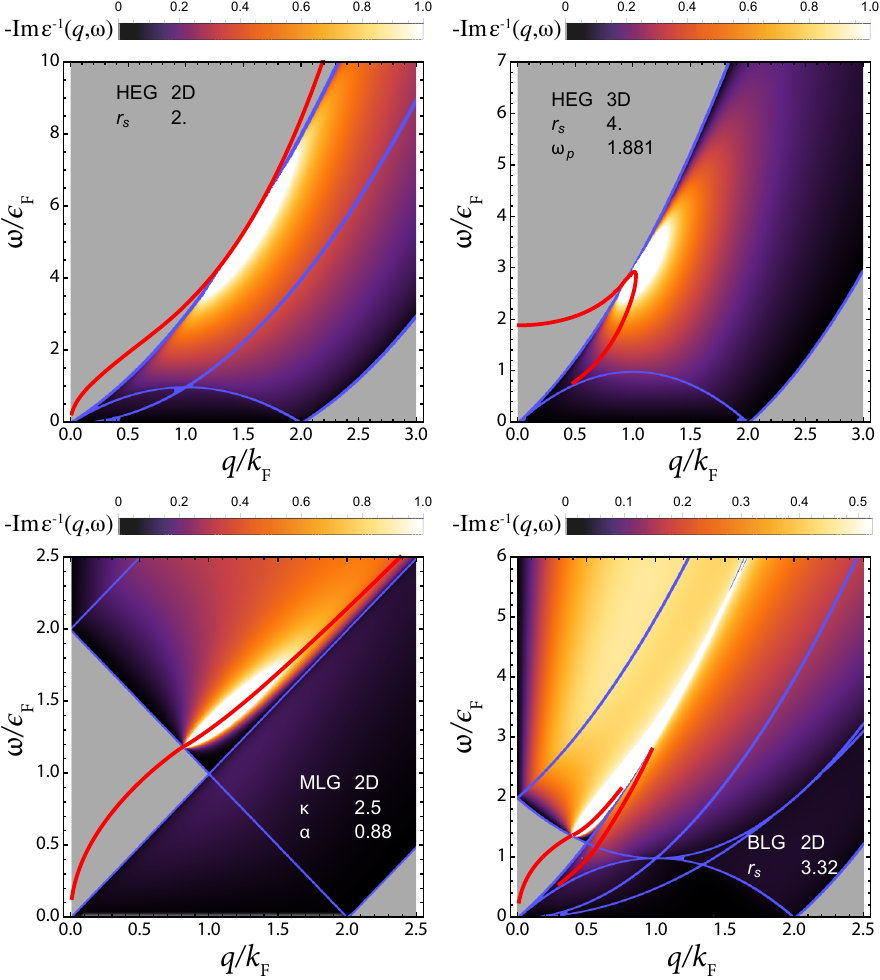}
\caption{\label{fig:im:eps} Imaginary part of the inverse dielectric function for the four
  studied systems. Red lines indicate the collective plasmonic mode. Blue lines separate
  different domains in the definition of polarizability such as shown in
  Fig.~\ref{fig:chi:dom} and indicate removable singularities.}
\end{figure}

The random phase approximation for the inverse dielectric function, $\im
\varepsilon_0(q,\w)^{-1}$ is a very important ingredient of the subsequent correlated
calculations because this gives (up to the Coulomb prefactor $v(q)$) the spectral function
of the screened interaction (Appendix~\ref{sec:app:prop}). A general overview of this
quantity is shown in Fig.~\ref{fig:im:eps}. It is very fortunate that for all four studied
systems it can be found in analytic form facilitating numerical calculations. Below, we
collect all needed formulas and additionally present the exchange self-energy, which
enters Eq.~\eqref{eq:re:sgm:R}.

In the following we express the electron density $n$, which is the central control
parameter, in SI units in order to make a connection with experiment.  All other
quantities are expressed in atomic units. Some simplification of formulas is possible to
achieve by rescaling momenta and energies by the Fermi momentum $\kF$ and energy $\eF$,
respectively. This will be explicitly indicated.
\subsection{2D HEG}
This is probably the best studied many-body system~\cite{czachor_dynamical_1982,santoro_electron_1989}. There is
only one relevant parameter\,---\,the Wigner-Seitz radius $r_s$. It is given in terms of
electronic density $n$ as follows:
\begin{align}
  a_\text{B}r_s&=\left(\frac{1}{\pi n}\right)^{1/2}.
\end{align}
In the case of systems with an effective electron mass $m_0$ and a background dielectric
constant $\kappa=4\pi\varepsilon\varepsilon_0$, one can redefine the Bohr radius as
\begin{align}
  \tilde{a}_\text{B}&=\frac{\kappa\hbar^2}{m_0e^2}
\end{align}
and still have the same relation between the density and $r_s$.  The Coulomb potential
$v(q)$, the Fermi momentum $\kF$ and the density of states at the Fermi energy
$\rho_\sigma(\eF)$ in atomic units read
\begin{align}
  v(q)&=\frac{2\pi}{q},&
  \kF&=\frac1{\alpha_2 r_s },&
  \rho_\sigma(\eF)&=\frac{1}{2\pi},
  \label{eq:par:heg:2d}
\end{align}
where we additionally defined the constant
\begin{align}
  \alpha_{2}&=1/\sqrt{2}.
\end{align}
Introducing scaled variables
\begin{align}
  y&=q/\kF,& \xi&=\w/\eF,
  \label{eq:scaling}
\end{align}
the dielectric function $\enR(k,\omega)\equiv\enR(y,\xi)$
reads~\cite{stern_polarizability_1967}
\begin{align}
  \re\enR(y,\xi)&=1+\frac{2\alpha_2 r_s}{y^2}\left[y+f_r\mleft(\nicefrac{\xi}{y}-y\mright)-f_r\mleft(\nicefrac{\xi}{y}+y\mright)
  \right],\label{eq:re:eps:heg:2d}\\
  \im \enR(y,\xi)&=\frac{2\alpha_2 r_s}{y^2}\left[f_i\mleft(\nicefrac{\xi}{y}-y\mright)
  -f_i\mleft(\nicefrac{\xi}{y}+y\mright)\right],
  \label{eq:im:eps:heg:2d}
\end{align}
with
\begin{align*}
  f_r(z)&=\sgn(z)\thf(\nicefrac14z^2-1)\sqrt{\nicefrac14z^2-1},\\
  f_i(z)&=\thf(1-\nicefrac14z^2)\sqrt{1-\nicefrac14z^2}.
\end{align*}  
For small momentum values, the expression in brackets of Eq.~\eqref{eq:re:eps:heg:2d}
suffers from the precision loss. Therefore, in this limit the approximate formula 
\begin{align}
  \re\enR(y,\xi)&=1-\frac{qy}{\xi^2}, &q=4 \alpha_2 r_s.
\end{align}
should be used entailing the small-momenta plasmon dispersion $\Omega(y)\approx\sqrt{yq}$.
It can also be found analytically (see Eq.\,5.54 of Ref.~\cite{giuliani_quantum_2005}):
\begin{align}
  \Omega(y)&=\frac{\sqrt{y} (2 y+q) \sqrt{y^4+y^3 q+q^2}}{q \sqrt{y+q}}.
\end{align}
The critical wave-vector does not have a nice analytical expression. However, one can show
that $k_c\sim\sqrt{q}$. Notice that even though $ \Omega(y)>y(y+2)$ for $y>k_c$ there is
no plasmon above the critical vector because in reality the plasmon becomes damped by
entering the continuum, where the above solution is not valid.

The $f$-sum rule reads in rescaled units
\begin{align}
  -\frac2\pi\int_{0}^\infty\dd  \xi\,\xi\im\mleft[\frac{1}{\eR(y,\xi)}\mright]&=\alpha_2 r_s^3 y.
\end{align}

The exchange part of the electron self-energy,
\begin{align}
  \Sigma_x(k)&=-\int_{|q|<\kF}\frac{\dd^2 q}{(2\pi)^2}\frac{2\pi}{\kappa|\vec k-\vec q|},
\end{align}
can only be expressed~\cite{giuliani_quantum_2005} in terms of the \emph{complete elliptic
  integrals} (see Sec.\,8.\,112 in Ref.~\cite{gradshteyn_table_2007}).
\begin{align}
  \bar\Sigma_x(y)&=-\frac{2}{\pi} \alpha_2 r_s f_{2D}(y),\qquad \bar\Sigma_x=\Sigma_x/\eF,\\
  f_{2D}(y)&=\begin{cases}
  E(y),&y\le1,\\
  y\left[E\mleft(\tfrac{1}{y}\mright)-\left(1-\tfrac{1}{y^2}\right)K\mleft(\tfrac1{y}\mright)\right],&y>1.
  \end{cases}
  \label{eq:f:2d}
\end{align}
\subsection{3D HEG}
This system also depends on a single parameter\,---\,the Wigner-Seitz radius
\begin{align}
  a_\text{B}r_s&=\left(\frac{3}{4\pi n}\right)^{1/3}.
\end{align}
It has also been broadly studied~\cite{lundqvist_single-particle_1968}. The Coulomb
potential $v(q)$, the Fermi momentum $\kF$ and the density of states at the Fermi energy
$\rho_\sigma(\eF)$ read in atomic units
\begin{align}
  v(q)&=\frac{4\pi}{q^2},&
  \kF&=\frac1{\alpha_3 r_s},&
  \rho_\sigma(\eF)&=\frac{1}{2\pi^2\alpha_3 r_s},
  \label{eq:par:heg:3d}
\end{align}
where the relevant constant is defined as
\begin{align}
\alpha_3&=\left(\frac{4}{9\pi}\right)^{1/3}.
\end{align}
The dielectric function is (the Lindhard result)
\begin{align}
  \re\enR(y,\xi)&=1+\frac{\alpha_3 r_s}{\pi y^3}\left[
  2y +f_r(\nicefrac{\xi}{y}-y)-f_r(\nicefrac{\xi}{y}+y)
  \right],\\
\im \enR(y,\xi)&=\frac{\alpha_3 r_s}{y^3}
 \left[f_i\mleft(\nicefrac{\xi}{y}-y\mright)
  -f_i\mleft(\nicefrac{\xi}{y}+y\mright)\right],
\end{align}
with
\begin{align*}
  f_r(z)&=(1-\nicefrac14z^2)\log[(z+2)/(z-2)],\\
  f_i(z)&=\thf(\nicefrac14z^2-1)(1-\nicefrac14z^2).
\end{align*}
Notice a strong resemblance between the dielectric function in 2D and 3D. This is due to
the fact that upper and lower continuum frequencies are the relevant parameters in both
cases (Fig.~\ref{fig:chi:dom}). The shape of continuum is more complicated for MLG and
BLG. However, we will see below that they likewise enter expressions for
$\enR(y,\xi)$.

The $f$-sum rule is particularly simple in 3D systems. This is due to the form of the
Coulomb interaction proportional to $q^{-2}$~\eqref{eq:par:heg:3d}.  Rescaling the
frequency and momentum in the usual way~\eqref{eq:scaling} we get
\begin{align}
  -\frac2\pi\int_{0}^\infty\dd  \xi\,\xi\im\mleft[\frac{1}{\eR(y,\xi)}\mright]&=\w_p^2,
\end{align}
with the classical plasmon frequency ($\eF$ units)
\begin{align}
  \w_p&=4\sqrt{\frac{\alpha_3 r_s}{3\pi}}.
\end{align}
The exchange part of the electron self-energy reads
\begin{align}
  \Sigma_x(k)&=-\int_{|q|<\kF}\frac{\dd^3 q}{(2\pi)^3}\frac{4\pi}{|\vec k-\vec q|^2}.
\end{align}
Analytical expressions are well-known~\cite{giuliani_quantum_2005}
\begin{align}
  \bar\Sigma_x(y)&=-\frac{4}{\pi}\alpha_3 r_s f_{3D}(y),\qquad \bar\Sigma_x=\Sigma_x/\eF,\\
  f_{3D}(y)&=\frac{1}{2}+\frac{1-y^2}{4y}\log\abs*{\frac{1+y}{1-y}}.
\end{align}
\subsection{2D MLG}
\begin{figure}[] 
  \centering
    \includegraphics[width=\columnwidth]{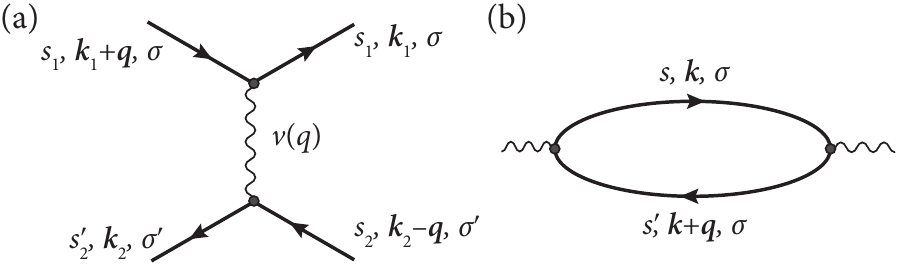}
\caption{\label{fig:v:mlg} Diagrammatic representation of (a) the Coulomb
  interaction~\eqref{eq:MLG:v} and (b) the polarization bubble in MLG. The latter
  illustrates the appearance of the spin $g_s$ and band $g_v$ degeneracy prefactors in the
  dielectric function, Eqs.~(\ref{eq:eps:mlg:re},\ref{eq:eps:mlg:im}). The overlap matrix
  elements~\eqref{eq:overlap} are represented by shaded vertices.}
\end{figure}

In the model approach to graphene, electronic states of the $\pi$-bands near a $K$ point
of the Brillouin zone are described by the $\bvec{k}\cdot \bvec{p}$ equation $\hat{\mH}_0
\bvec{F}(\bvec r)=\epsilon
\bvec{F}(\bvec{r})$~\cite{ando_screening_2006,das_sarma_electronic_2011}, where the
Hamiltonian reads
\begin{align}
  \hat{\mH}_0&=\vF
  \begin{pmatrix}0&\hat{p}_x-\ii \hat{p}_y\\
    \hat{p}_x+\ii \hat{p}_y&0
    \end{pmatrix}
  =\vF(\sigma_x \hat{p}_x+\sigma_y \hat{p}_y),  \label{eq:h0}
\end{align}
with $\hat{\bvec{p}}=(\hat{p}_x,\hat{p}_y)$ being the momentum operator and $\vF$ the Fermi
velocity (can be expressed in terms of the hopping integral and the lattice
constant~\cite{basov_colloquium:_2014}, the typically adopted value is
$10^6\,\text{m}/\text{s}=1/2.188\,\text{a.u.}$). The wave-function is then
\begin{align}
  \bvec{F}_{s,\bvec{k}}(\bvec
  r)&=|s,\bvec{k}\rangle\frac{1}{L} e^{\ii \bvec{k}\cdot\bvec{r}},&
  |s,\bvec{k}\rangle&=\frac1{\sqrt{2}}
  \begin{pmatrix}e^{-\ii \theta_{k}}\\s\end{pmatrix},\label{eq:bloch:wave}
\end{align}
where $L^2$ is the area of the system, and $k_x=k\cos\theta_k$, $k_y=k\sin\theta_k$,
$k=|\bvec k|$. The corresponding energy dispersion reads
\begin{align}
  \epsilon(k)&=s\vF k,\label{eq:MLG:disp}
\end{align}
which is different from previous cases in two important ways: i) the well-known linear
momentum-dependence and ii) the presence of two bands indicated by the band index $s=\pm1$
and, as a consequence, the presence of additional matrix elements in the Coulomb operator
(Fig.~\ref{fig:v:mlg})
\begin{align}
  \widehat{V}&=\frac{1}{2L^2}\sum_{\bvec{q},\bvec{k}_1,\bvec{k}_2}\sum_{s_1,s_2,s_1',s_2'}
  \langle s_1',\bvec{k}_1+\bvec{q}| s_1,\bvec{k}_1\rangle
  \langle s_2,\bvec{k}_2-\bvec{q}|s_2',\bvec{k}_2\rangle\nn\\
  &\qquad\times\sum_{\sigma,\sigma'}\frac{2\pi}{\kappa q}
  \hat c_{s_1',\bvec{k}_1+\bvec{q},\sigma}^\dagger\hat c^\dagger_{s_2,\bvec{k}_2-\bvec{q},\sigma'}
  \hat c_{s_2',\bvec{k}_2,\sigma'} \hat c_{s_1,\bvec{k}_1,\sigma}.
  \label{eq:MLG:v}
\end{align}
Thus, basis functions are labeled by the momentum $\bvec k$, band index $s$ and spin
$\sigma$. In view of the dispersion~\eqref{eq:MLG:disp}, the noninteracting GF is diagonal
in $s$ and $\sigma$. We are interested in the electron self-energy diagonal in the band
indices. Furthermore, the calculations are typically performed at finite doping (extrinsic
graphene) and with dielectric function modified by the presence of substrate. We will
focus on the SiO$_2$ substrate $\kappa=(1+\varepsilon_{\text{SiO}_2})/2=2.45$, consider
the case of the \emph{electron doping}, i.e., that the Fermi level is above the Dirac
point, and follow the notations from the previous sections that
\begin{align}
  \kappa&=4\pi\varepsilon\varepsilon_0,
\end{align}
with $\varepsilon_0$ being the vacuum electric permittivity.  The Fermi momentum and
energy depend on the square root of the electron density
\begin{align}
  \kF&=\aB\sqrt{\frac{4\pi n}{g}}=\frac{1}{\langle r\rangle},&\eF&=\vF\kF,
  \label{eq:MLG:kF}
\end{align}
$g=g_sg_v$ where $g_s=2$ is the spin and $g_v=2$ is the valley degeneracy,
respectively. $\langle r\rangle$ is the averaged inter-electron distance.

The parameter characterizing the level of correlations in the system is given by the ratio
of the Coulomb $E_C$ and the kinetic $E_K$ energies, as is therefore the counterpart of
$r_s$ for the homogeneous electron gas
\begin{align}
  \alpha&=\frac{E_C}{E_K}=\frac{1}{\kappa\,\langle r\rangle} \frac{1}{\vF \kF}
  =\frac{1}{\kappa \vF}\simeq \frac{2.188}{\kappa},
  \label{eq:MLG:alpha:val}
\end{align}
for instance, $\alpha=2.188$ for MLG in vacuum and $\alpha=0.875$ for the SiO$_2$
substrate.  The dielectric function has been computed by Hwang and Das
Sarma~\cite{hwang_dielectric_2007} and by Wunsch \emph{et
  al.}~\cite{wunsch_dynamical_2006}. We will use the latter form:
\begin{align}
  \re \enR(y,\xi)&=1+\frac{g\alpha}{y}
  +g\alpha f(y,\xi) G_r(y,\xi),  \label{eq:eps:mlg:re}\\
  \im \enR(y,\xi)&=g\alpha f(y,\xi) G_i(y,\xi),\\
  f(y,\xi)&=\frac{1}{8}\frac{y}{\sqrt{|\xi^2-y^2|}},
  \label{eq:eps:mlg:im}
\end{align}
Functions $G_r$ and $G_i$ are defined in Appendix~\ref{app:mlg}. The density of states at the
Fermi level reads
\begin{align}
  \rho_{\sigma,s}&=(2\pi \vF)^{-1}.
\end{align}
The static polarizability normalized at this number is plotted in
Fig~\ref{fig:chi:k:fsum}(a).  Due to the presence of infinite sea of electrons below the
Dirac point, the $f$-sum rule diverges as demonstrated by Hwang, Throckmorton, and Das
Sarma~\cite{hwang_plasmon-pole_2018}, the integrand of the $f$-sum is illustrated in
Fig.~\ref{fig:chi:k:fsum}(b).

Due to this fact, a momentum cut-off $k_c$ needs to be introduced for the momentum
integrals. In realistic system this is not a problem because of the bands flattening due
to lattice effects~\cite{trevisanutto_ab_2008}.  For the idealistic model that we consider
here, $k_c$ is an explicit parameter of the theory. We adopt
\begin{align}
  k_c&=y_c\kF=10\kF.\label{eq:kc}
\end{align}
The exchange self-energy can be written in the form 
\begin{align}
  \Sigma_{x,s}(k)&=-\!\sum_{s'=\pm1}\!\int\!\!\frac{\dd^2q}{(2\pi)^2}\nFs{s'}(\vec
  k-\vec q) \frac{2\pi}{\kappa q}F_{s,s'}(\vec k,\vec k-\vec q).
  \label{eq:Sx:mlg:def}
\end{align}
In this equation, $F_{s_1,s_2}(\bvec k_1,\bvec k_2)$ takes into account the probability for
an electron with momentum $\bvec k_1$ in the band $s_1$ to scatter into the state with
momentum $\bvec k_2$ in the band $s_2$. It depends on the relative angle $\theta_{12}$
between the two momenta,
\begin{align}
   \langle s_2,\bvec k_2|s_1,\bvec k_1\rangle&=\tfrac12(1+s_1s_2e^{\ii \theta_{12}}),\label{eq:overlap}\\
   F_{s_1,s_2}(\bvec k_1,\bvec k_2)=|\langle s_2,\bvec k_2|s_1,\bvec k_1\rangle|^2&
   =\tfrac12(1+s_1s_2\cos\theta_{12}). \label{eq:F12}
\end{align}
Using the Fermi energy and momentum units, dividing into intrinsic (present in pristine
graphene) and extrinsic (due to carriers injection by doping or gating) contributions,
shifting by the constant so that the self-energy is zero at the Dirac point
($\Sigma_{x,\pm1}(0)=0$) we obtain for Eq.~\eqref{eq:Sx:mlg:def}
\begin{align}
  \bar\Sigma_{x,s}(y)&=\bar\Sigma_{x,s}^\text{int}(y)+\bar\Sigma_{x,s}^\text{ext}(y)+\frac{\alpha}{2}(1+y_c),
\end{align}
with
 \begin{align}
   \bar\Sigma_{x,s}^\text{int}(y)&=-\frac{\alpha y_c}{\pi}\left[\frac{\pi}{2}-sg\mleft(\frac{y}{y_c}\mright)\right],\\
  \bar\Sigma_{x,s}^\text{ext}(y)&=-\frac{\alpha}{\pi}\Big[f_{2D}(y)+ s h(y)\Big].
  \label{eq:Sx:mlg}
 \end{align}
The function $f_{2D}(y)$ has already been defined for 2D HEG~\eqref{eq:f:2d}. The former
intrinsic part results from the integration over the $s'=-1$ band from zero to the
momentum cut-off $k_c=y_c \kF$.  Functions $h(y)$ and $g(y)$ have a representation
(correcting $g(y)$ in the original derivation by Hwang, Hu and Das
Sarma~\cite{hwang_density_2007}):
\begin{align}
  h(y)&=y\begin{dcases} \tfrac{\pi}{4}\log\tfrac{4}{ye^{1/2}}-\int_{0}^y \!\tfrac{\dd x}{x^{3}}
  \left[K(x)-E(x)-\tfrac{\pi x^2}{4}\right]&y\le1,\\
  \int_{0}^{1/y} \!\dd x\,[K(x)-E(x)]&y>1;
  \end{dcases}\\
  g(x)&=\frac{1}{4}\int_0^{1}\!\dd y \int_0^{2\pi}\! \dd\theta \frac{x-y\cos \theta}{\sqrt{x^2+y^2-2xy\cos\theta}}\nn\\
  &\qquad=\frac{\pi}{2} \re\mleft\{  _3F_2\mleft(-\tfrac{1}{2},\tfrac{1}{2},\tfrac{1}{2};1,\tfrac{3}{2};\tfrac{1}{x^2}\mright)\mright\}.
\end{align}

\subsection{2D BLG}
Consider now two parabolic energy bands
\begin{align}
  \epsilon(k)&=sk^2/(2m_0).
\end{align}
Unlike MLG, the dispersion is an idealization of several materials with different number
of valleys $g_v$ and with large flexibility in the properties control with the help of
doping and the background dielectric constant. Here we focus on the $g_v=2$ case pertinent
to the bilayer graphene (a minimal two-band model for the Bernal $AB$
stacking~\cite{kotov_electron-electron_2012}). We have the following relations determining
the Fermi energy and momentum, and the Wigner-Seitz radius~\cite{sensarma_dynamic_2010}
\begin{align}
  \frac{\kF}{\aB}&=\sqrt{\frac{4\pi n}{g_sg_v}},&\eF&=\frac{\kF^2}{2m_0},&
  r_s&=\frac{g m_0}{\kappa \kF},&
  \rho_{\sigma,s}(\eF)&=\frac{m_0}{2\pi}.
\end{align}
One may also define the Wigner-Seitz radius as the ratio of two energies
\begin{align*}
   \tilde{r}_s&=\frac{E_C}{E_K}=\frac{1}{\kappa \langle r\rangle}\frac{2m_0}{\kF^2}
   =\frac{m_0}{\kappa}\frac{g_sg_v \aB}{(4\pi n)^{1/2}},&
   \aB\langle r\rangle=\left(\frac{1}{\pi n}\right)^{1/2}.
\end{align*}

Sensarma, Hwang and Das Sarma~\cite{sensarma_dynamic_2010} derived the polarizability of
this system separating the intra(inter)-band contributions $\Pi=\Pi_1+\Pi_2$. The former
originates from the intraband ($s=s'$) and the latter from interband ($s=-s'$)
transitions, Fig.~\ref{fig:v:mlg}(b),
\begin{align}
  \Pi_1(y,\xi)&=\frac1{\pi}\int_0^1\!\dd x\int_{-\pi}^{\pi}\!\dd \phi\,
  \frac{x }{\xi+\ii\eta-2 x y \cos (\phi )-y^2}\nn\\
  &\quad\times\left[1-\frac{y^2 \sin ^2(\phi )}{x^2+2 x y \cos (\phi )+y^2}\right],\\
  \Pi_2(y,\xi)&=-\frac1{\pi}\int_0^1\!\dd x\int_{-\pi}^{\pi}\!\dd \phi\,
  \frac{x}{\xi+\ii\eta+2 x^2+2 x y \cos (\phi )+y^2}\nn\\
 &\quad \times \frac{y^2\sin ^2(\phi )}{x^2+2 x y \cos (\phi )+y^2}.
\end{align} 
The retarded polarizability is given
by
\begin{align}
  \re \mP(y,\xi)&=\re \Pi(y,\xi)+\re \Pi(y,-\xi),\\
  \im \mP(y,\xi)&=\im\Pi(y,\xi)-\im\Pi(y,-\xi).
\end{align}
$\Pi_1$ is fully defined in the paper~\cite{sensarma_dynamic_2010}. There are, however,
some misprints in the extrinsic part that are corrected here in Appendix~\ref{app:blg}.  The
dielectric function is given by
\begin{align}
  \varepsilon_0^{\mathrm{R}}(y,\xi)&=1-\frac{r_s}{y}\mP(y,\xi).
\end{align}
Th static polarizability was derived by Hwang and Das Sarma~\cite{hwang_screening_2008}
and is plotted here for comparison with other systems in Fig.~\ref{fig:chi:k:fsum}(a). The
$f$-sum rule diverges for this system for the same reasons as for MLG. The integrand of
the $f$-sum is illustrated in Fig.~\ref{fig:chi:k:fsum}(c).  The exchange self-energy is
the same as for MLG~\eqref{eq:Sx:mlg}.
\begin{figure}[] 
  \centering
    \includegraphics[width=\columnwidth]{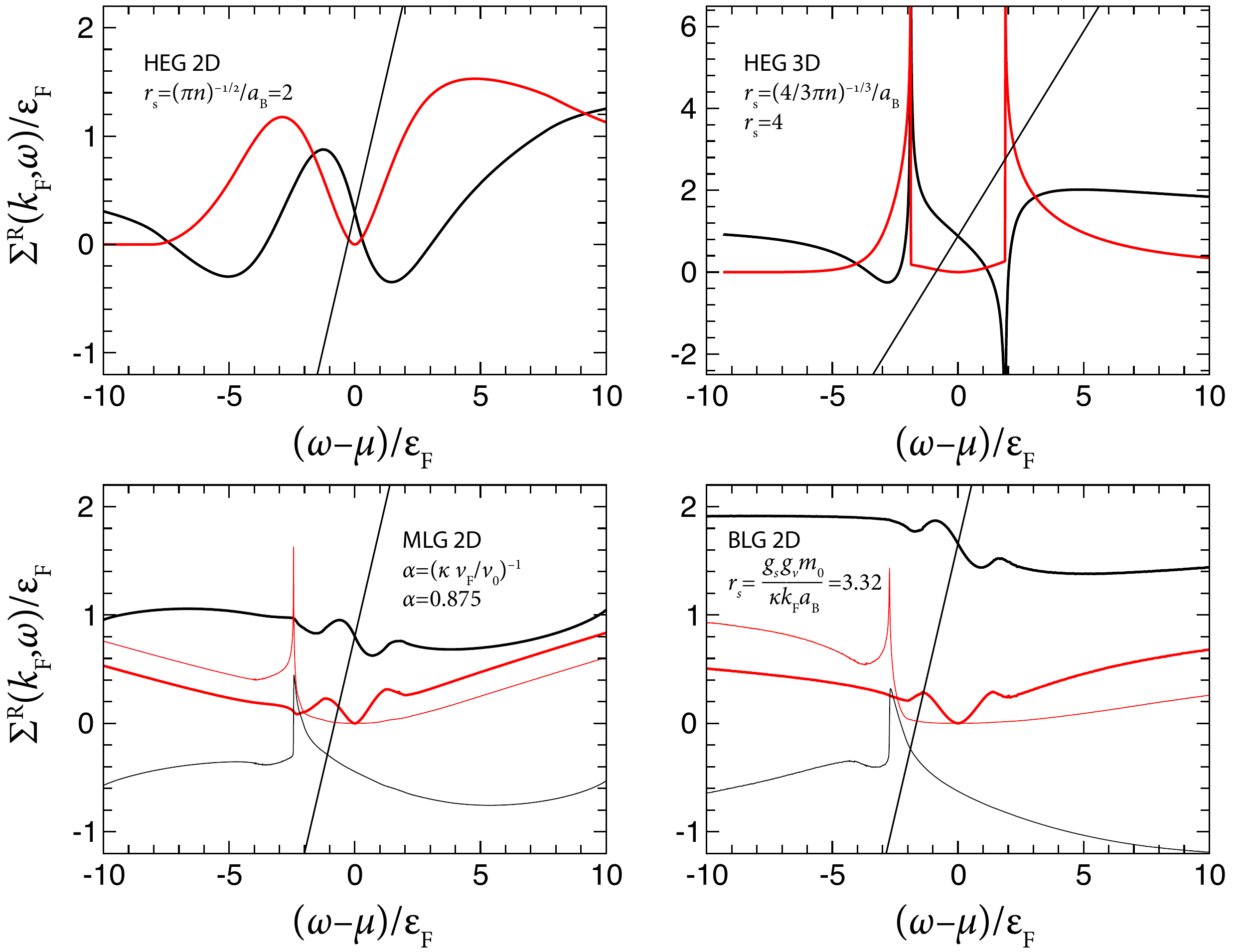}
\caption{\label{fig:sgm:kf} Solution of the Dyson equation for the Fermi level
  $k=\kF$. Intersection of straight line $y=\w-\eF-\Delta\mu$ and $y=\re\SgR(\kF,\w)$
  (thick black curve) yields the real part of quasiparticle energy. The chemical potential
  shift $\Delta \mu$ is selected (see Appendix~\ref{app:dyson}) as to have the imaginary
  part zero in accordance with the Fermi liquid assumption. Red curves stand for
  $\nicefrac12\Gamma(\kF,\w)=-\im\SgR(\kF,\w)$.  In the case of MLG and BLG thick/thin
  curves correspond to $\Sigma^\mathrm{R}_{s=\pm1}(\kF,\w)$.}
\end{figure}
\subsection{$G_0W_0$ calculations}
\begin{figure}[] 
  \centering
    \includegraphics[width=\columnwidth]{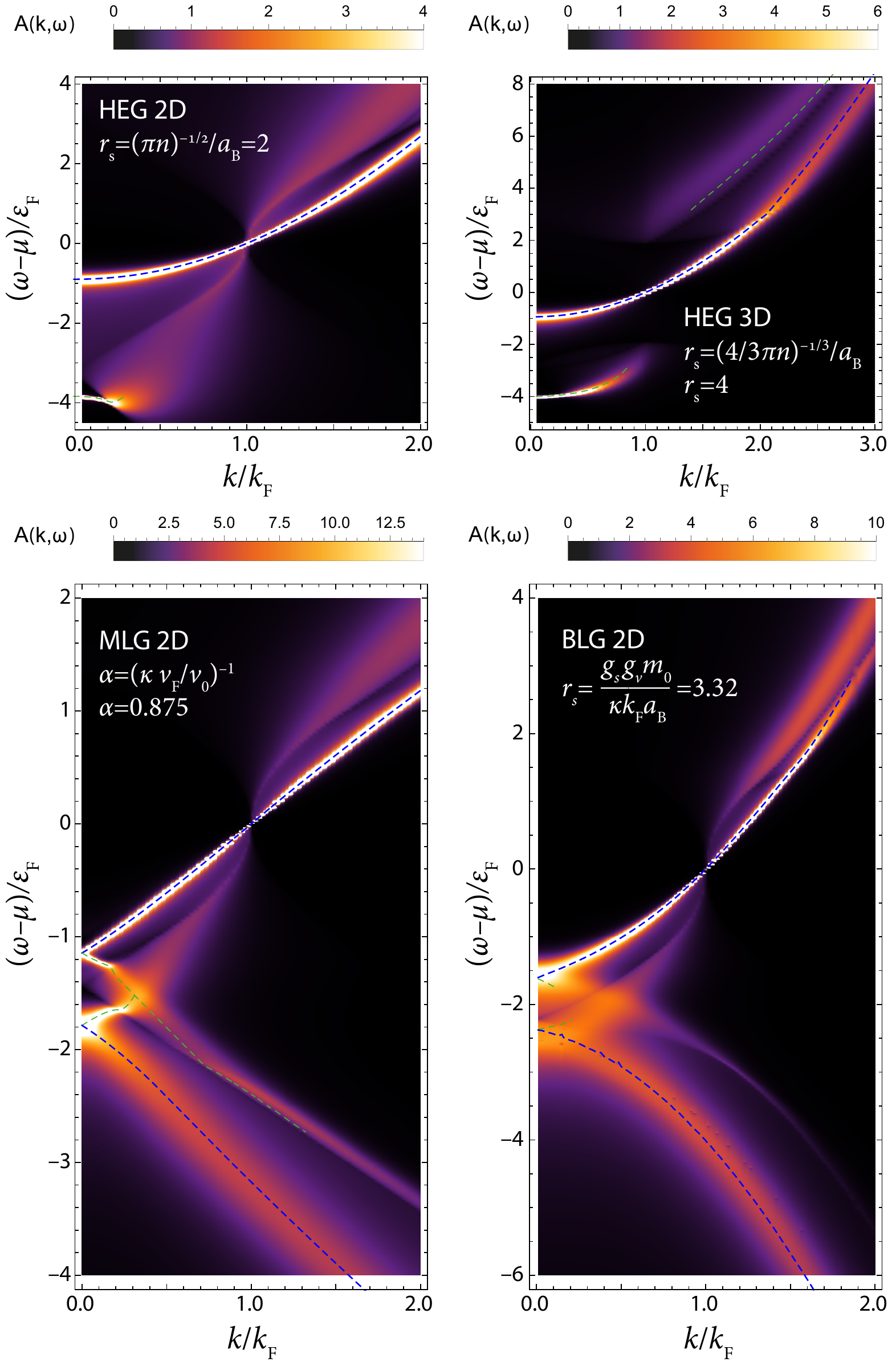}
\caption{\label{fig:A:gw} The electron spectral function from the solution of the Dyson
  equation with self-energy in $G_0W_0$ approximation. For MLG and BLG, the spectral
  function is a matrix in the band indices, the trace of it is shown here.}
\end{figure}
Before presenting calculations with vertex functions, we overview the electron self-energy
in the simplest $G_0W_0$ approximation, Eq.~\eqref{eq:sgm:aa}. The self-energy is depicted
in Fig.~\ref{fig:sgm:kf}, the respective electron spectral function
\begin{align}
  A(k,\w)&=-2\im G^\text{R}(k,\w)
\end{align}
is shown in Fig.~\ref{fig:A:gw}.
\paragraph{HEG} systems have a long history of studies:
Lundqvist~\cite{lundqvist_single-particle_1968}, Hedin~\cite{hedin_new_1965} and
self-consistent $GW$ calculations by von Barth and
Holm~\cite{von_barth_self-consistent_1996,holm_fully_1998} in 3D and Giuliani and
Quinn~\cite{giuliani_lifetime_1982}, Santoro and Giuliani~\cite{santoro_electron_1989},
Zhang and Das Sarma~\cite{zhang_quasiparticle_2005}, Lischner \emph{et
  al.}~\cite{lischner_satellite_2014} in 2D. Because of the way how 2D HEG is engineered
(its properties can be tuned by doping, the electron concentration), it is easy to go to
strongly correlated regime and still have a homogeneous system. Therefore, correlations
beyond $G_0W_0$ have been included almost from the beginning. Thus, Santoro and Giuliani
included the many-body local fields $G_{\pm}$ in the calculation of screening and employed
the plasmon-pole approximation. This yields the self-energy resembling the 3D case and
results in a more pronounced plasmon peak as compared to the $G_0W_0$ calculations.

\paragraph{Graphene}
There are two peculiarities in the case of MLG and BLG systems. (i) The electron
dispersion and self-energies additionally carry the valley index $s$ resulting in the
following modification of Eq.~\eqref{eq:sgm:aa}:
\begin{align}
  \Sigma_{GW,s}^{<}(k,\w)&=2\ii\pi\sum_{s'=\pm1}\int_0^{q_{1}\le k_c}
  \frac{\dd^2q_{1}}{(2\pi)^2}  \int_0^\infty\! \dd\nu_1\,
   F_{s,s'}(\vec{k};\vec{k}_{1})\nn\\ &\qquad\times
   \nFs{s'}(k_{1}) C(q_{1},\nu_1)\dlf(\w+\nu_1-\epsilon_1),
  \label{eq:sgm:aa:gr}
\end{align}
where the scattering matrix element $F_{s,s'}$ is given by Eq.~\eqref{eq:F12}. (ii) Due to
the presence of an infinite electron sea below the Dirac point the diverging momentum
integrals need to be regularized with the help of cut-off~\eqref{eq:kc}. One can also
introduce a frequency cut-off without compromising the accuracy.

Hwang and Das Sarma~\cite{hwang_quasiparticle_2008} and Polini \emph{et
  al.}~\cite{polini_plasmons_2008} performed calculations for MLG, more extensive
investigations for a range of momenta are in Refs.~\cite{bostwick_observation_2010,
  walter_effective_2011,carbotte_emergence_2012,das_sarma_velocity_2013}.
Respective calculations for BLG have been performed by Sensarma, Hwang and Das
Sarma~\cite{sensarma_quasiparticles_2011} and Sabashvili \emph{et
  al.}~\cite{sabashvili_bilayer_2013}.

%=============================================================
%            SIGMA aA (second-order exchange)
%=============================================================
\section{$\Sigma_{a\bar{a}}$: scattering accompanied by the generation of a $ph$-pair
  with exchange \label{sec:SGM:aA}} $\Sigma_{a\bar{a}}$ is the main objective of this
work. It describes the simplest second-order process in which a particle scatters giving
rise to an additional particle-hole pair in the final state, Fig.~\ref{fig:SGMaA}.  It is
obtained by gluing two half-diagrams $D^{(a)}$ with a permutation, and therefore does not
lead to a PSD spectral functions on its own. However, the inclusion of an unpermuted
configuration gives rise to $\Sigma_{aa}$ restoring the PSD property. In this work
$\Sigma_{a\bar{a}}$ is computed according to Eq.~\eqref{eq:sgm:aA}, which needs some
modifications in the case of graphene in order to account for the band indices. After
discussing this technical point in Sec.~\ref{sec:gph}, we consider the influence of
screening on $\Sigma_{a\bar{a}}$ in Sec.~\ref{sec:sgm2x:heg}, the cancellations between
$\Sigma_{aa}$ and $\Sigma_{a\bar{a}}$ in the asymptotic regime in Sec.~\ref{sec:asympt},
and finally focus on the resulting quasiparticle properties in Sec.~\ref{sec:qp}.

\subsection{Computation for MLG and BLG systems\label{sec:gph}}
In the case of graphene, Eq.~\eqref{eq:sgm:aA} additionally gets a sum over three internal
band indices and an additional factor, which is a product of the four wave-function
overlaps,
\begin{multline}
  F_{s_0,s_1s_2s_3}(\bvec k_0;\bvec k_1,\bvec k_2,\bvec k_3)=\langle s_0,\bvec k_0|s_1,\bvec k_1\rangle
  \langle s_1,\bvec k_1|s_3,\bvec k_3\rangle\\
  \times \langle s_3,\bvec k_3|s_2,\bvec k_2\rangle \langle s_2,\bvec k_2|s_0,\bvec k_0\rangle
  =\frac18\Big\{1+\sum_{i<j}s_is_j\cos\theta_{ij}\\
  +s_0s_1s_2s_3\cos (\theta_{01}+\theta_{32})\Big\},
  \label{eq:F:2x}
\end{multline}
where $\theta_{ij}$ is the angle between the respective momenta. The second-order exchange
then takes a form
\begin{multline}
  \Sigma_{2x,s}^<(k,\w)=-2\ii\pi\!\sum_{s_1,s_2,s_3} \!\iint\!\dd(\Omega_{1,2})\,
  F_{s,s_1s_2s_3}(\vec{k};\vec{k}_{1},\vec{k}_{2},\vec{k}_{3})\\
  v(q_{1}) v(q_{2}) \nF(\epsilon_1)\nF(\epsilon_2) \nbF(\epsilon_3)
  \delta(\w-\epsilon_1-\epsilon_2+\epsilon_3),
  \label{eq:sgm:2x}
\end{multline}
where the momenta are defined by Eq.~\eqref{eq:momentum:cons} and depicted in
Fig.~\ref{fig:diag2}(a). It should be noted that our original PSD construction was
formulated for the systems free of the ultra-violet divergences. Here it is applied
graphene, for which the momentum integrals are regularized with the wave-vector cutoff
$k_c$ with the justification that the regularization can be implemented on the level of
Hamiltonian.

\subsection{$\Sigma_{2x}$ results for 3D HEG\label{sec:sgm2x:heg}}
\begin{figure}[] 
  \centering \includegraphics[width=\columnwidth]{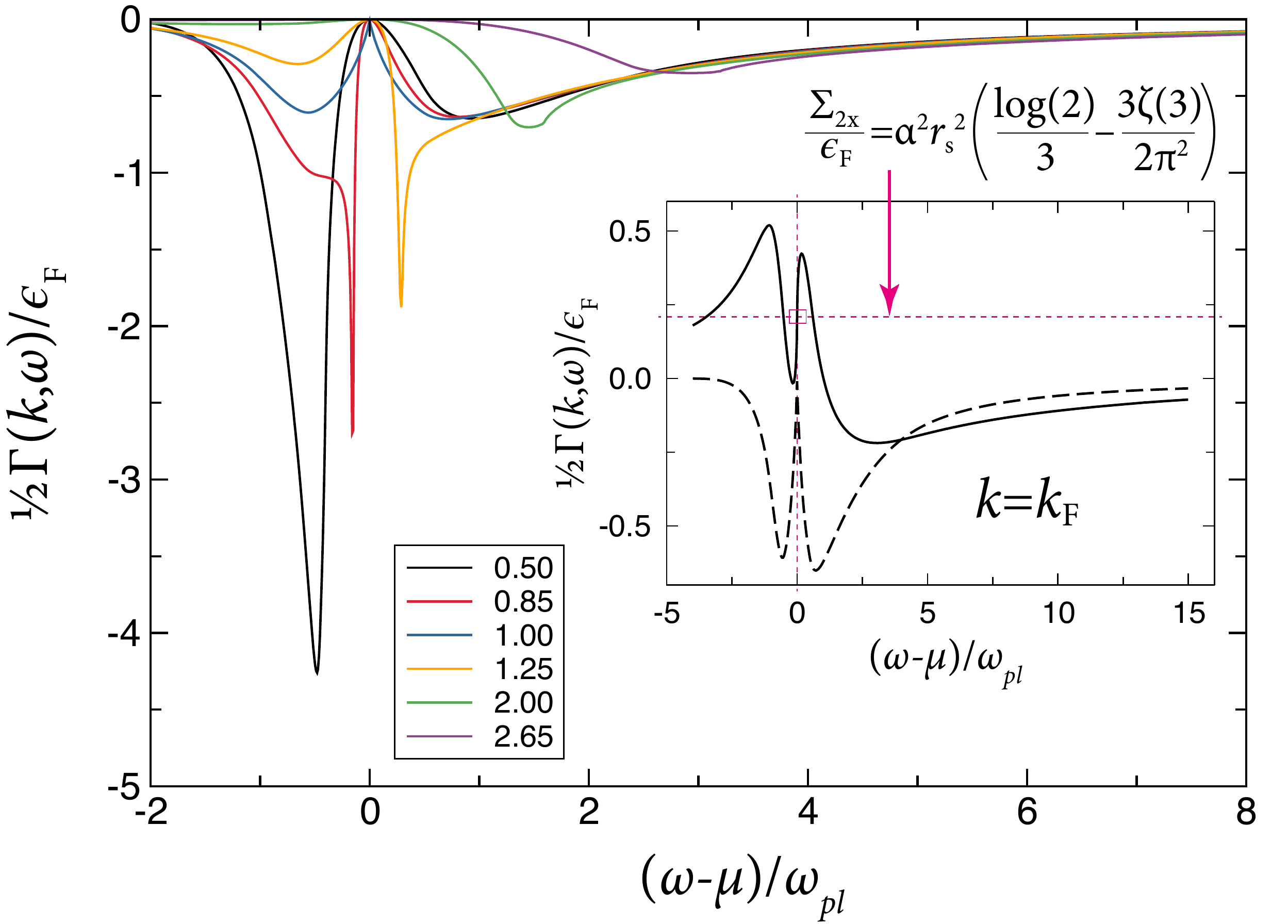}
\caption{\label{fig:SGMa_v} $\Sigma_{2x}$ at different momentum values. The inset shows
  the real self-energy part (solid line) obtained from the Hilbert transform of
  $\Gamma_{a\bar{a}}(\kF,\omega)$. Its on-shell value ($\epsilon_{2x}(\kF,\mu)=0.204976$)
  has only 2.4\% deviation from the indicated analytic result (0.210073).}
\end{figure}
Corresponding imaginary part is plotted in Fig.~\ref{fig:SGMa_v}. The Hilbert transform
(Appendix~\ref{sec:app:hilbert}) yields the real part.  On the inset we see a very good
agreement of our numerical result with the analytical expression
\begin{align*}
  \epsilon_{2x}=\re\Sigma_{2x}(\kF,\mu)/\eF=\frac{\alpha^2r_s^2}{2\pi^2}
  \left(\frac{2\pi^2}{3}\ln(2)-3\zeta(3)\right),
\end{align*}  
that is known due to the calculations of Glasser and Lamb~\cite{glasser_analysis_2007} and
Ziesche~\cite{ziesche_self-energy_2007} or from the second-order correction to the total
energy computed by Onsager \emph{et al.}~\cite{onsager_integrals_1966}. According to the
Hugenholtz-van Hove-Luttinger-Ward theorem they are equal. Despite
claims~\cite{isihara_exact_1980}, it seems impossible to get the respective expression in
analytic form for 2D HEG~\cite{angilella_second_2018}.

\begin{figure}[] 
  \centering
    \includegraphics[width=\columnwidth]{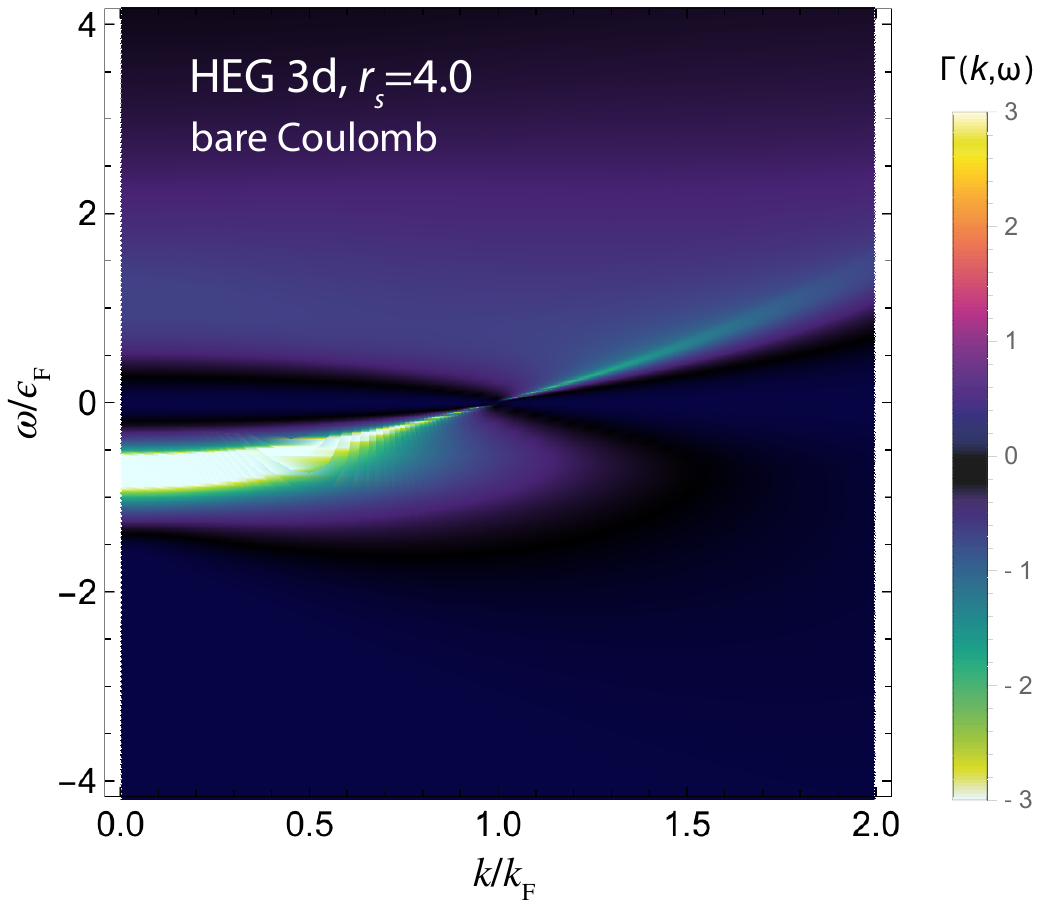}
\caption{\label{fig:SGMaA:kin} The momentum- and energy-resolved scattering rate computed
  with the bare Coulomb interaction for $r_s=4$. It is negative and possesses a strong
  peak due to the small-momentum forward scattering. Screening greatly reduces its
  magnitude, whereas $\Sigma_{aa}$ compensates for the negative values.}
\end{figure}

Going away from $k=\kF$, the self-energy first develops an additional sharp peak in the
vicinity of $\omega=\mu$ as seen for $k=0.85\kF, 1.25\kF$, which eventually becomes
smeared out, Fig.~\ref{fig:SGMaA:kin}. This is a rather disturbing fact because large
negative values need to be compensated by $\Sigma_{aa}$, which does not have any
singularities in this energy range. Thus, a better understanding of the origin of this
peak is needed.

\begin{figure}[] 
  \centering
\includegraphics[width=0.99\columnwidth]{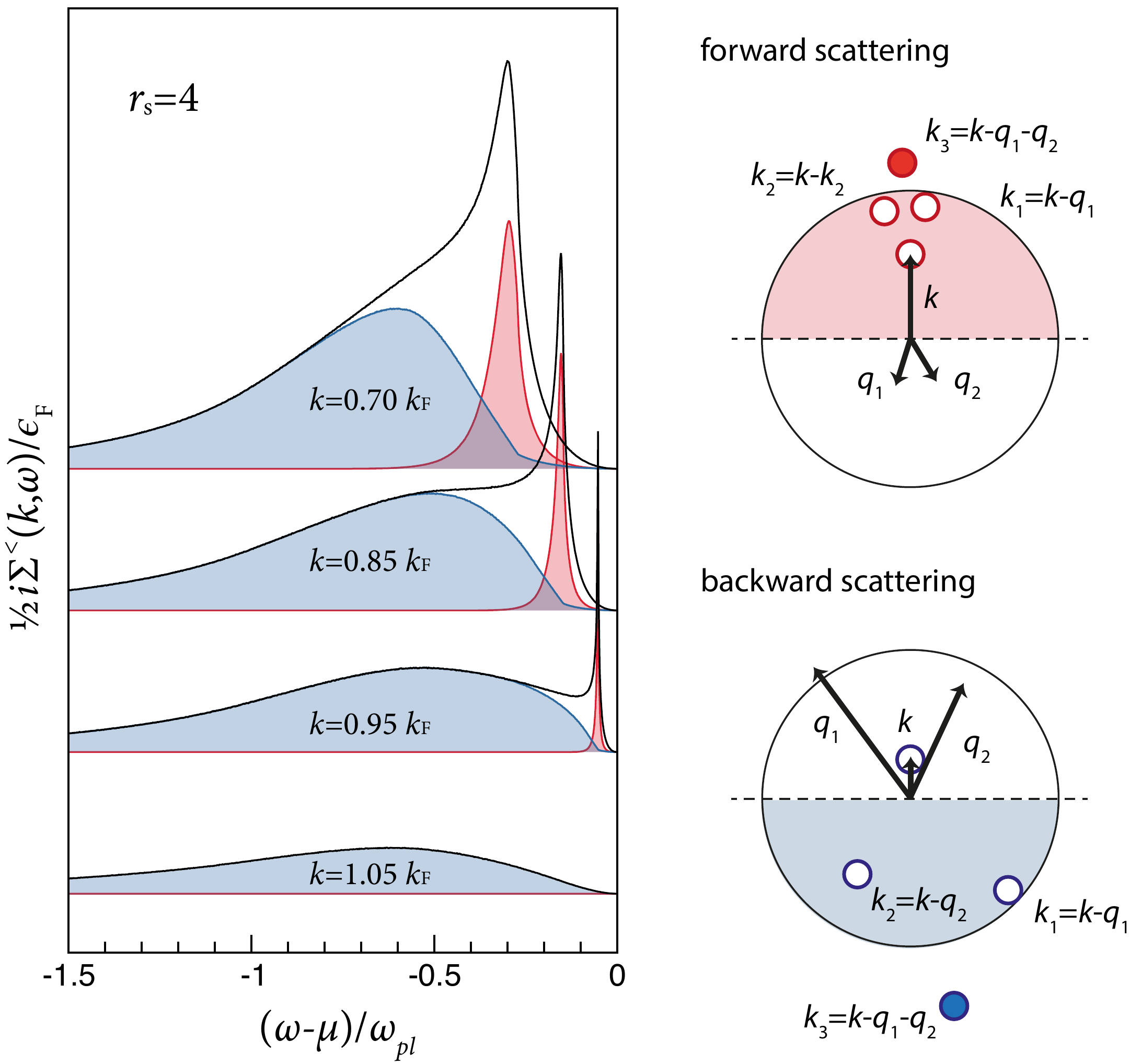}
\caption{\label{fig:SGMa_FwBc} $\Sigma_{2x}$ at different momentum values resolved with
  respect of forward (red) and backward (blue) scattering mechanisms. For a hole in the
  center of the Fermi sphere $k\ll \kF$, backward scattering dominates, while for a hole
  close to the Fermi surface $k\approx \kF$ forward scattering with a small momentum
  transfer gives rise to a sharp peak.}
\end{figure}

In Fig.~\ref{fig:SGMa_FwBc} we re-plot $i\Sigma_{a\bar{a}}^<$ computed with bare Coulomb
interaction for different momentum values paying attention to the kinematic aspects. In
particular, we are interested in the distribution of momenta carried by the two
interaction lines $\vec{q}_{1,2}$ and, correspondingly, in the configuration of the final
state formed by two holes with momentum $\vec{k}_{1,2}=\vec{k}-\vec{q}_{1,2}$ and a
particle $\vec{k}_{3}=\vec{k}-\vec{q}_{1}-\vec{q}_{2}$, Fig.~\ref{fig:SGMaA}.  It is, of
course, difficult to depict all the multitude of possibilities taking place in our Monte
Carlo simulations. However, a useful classification of the involved physical processes can
be found: we distinguish the forward and the backward scatterings scenarios. The former is
defined as a process in which $\vec{q}_{1}$ and $\vec{q}_{2}$ are \emph{anti-parallel} to
the initial hole momentum, i.\,e., the scalar products $(\vec{k},\vec{q}_{1})$ and
$(\vec{k},\vec{q}_{2})$ are negative. In this case the initial hole state with momentum
$k$ gets transformed into two-hole states with momenta in the same Fermi hemisphere
(red). For the backward scattering both of these products are positive and,
correspondingly, the final hole states are in the opposite hemisphere. From the scheme
depicted in Fig.~\ref{fig:SGMa_FwBc} it becomes evident that there is a very limited phase
space for the forward mechanism if the initial hole is in the vicinity of Fermi sphere,
$k\rightarrow \kF$. In order to guarantee that $k_{1,2}\le \kF$ and $k_{3}>\kF$ the
interaction momenta $\vec{q}_{1,2}$ must be small and almost collinear with $\vec{k}$. As
a result we have a ``hot-spot'' in the momentum space where all the permitted
configurations contribute in a very narrow energy interval giving rise to a pronounced
forward peak. If the initial state is closer to the center of Fermi sphere, there are less
restrictions on the possible scattering angles. Therefore, the forward peak broadens, and
for larger energy transfers the backward scattering dominates. It is interesting to notice
that the mixed mechanism, i.\,e., where one hole is in the forward and another in the
backward direction, has a rather small contribution and takes place at intermediate
energies.

\begin{figure}[] 
  \centering
\includegraphics[width=0.84\columnwidth]{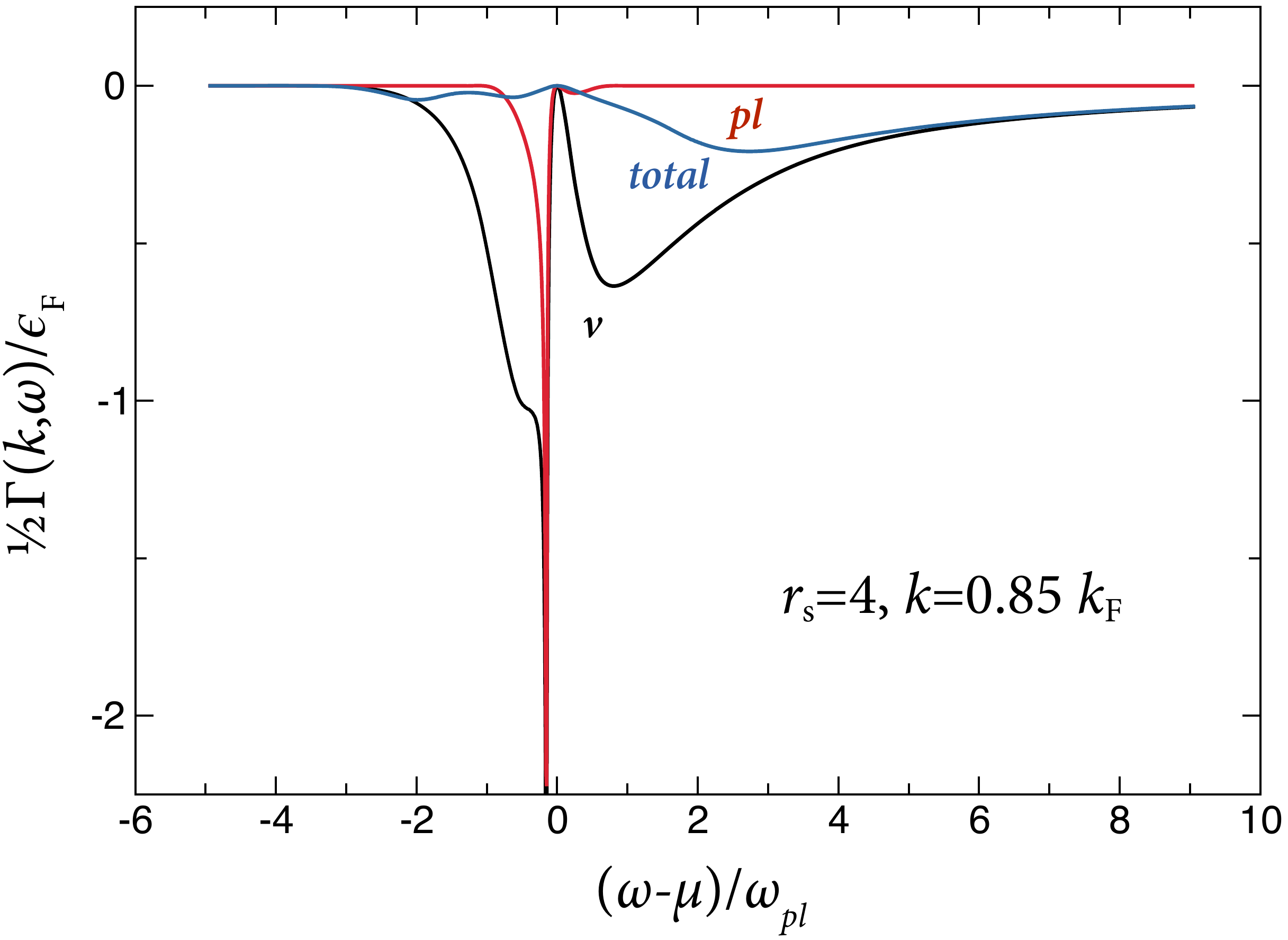}
\caption{\label{fig:SGMa_v_pl_tot} $\Sigma_{a\bar{a}}$ computed with full RPA screenings
  $W_0$ (total), plasmon pole approximation for $W_0$ ($pl$), and bare Coulomb interaction
  ($v$). Screening, which is operative at small momentum transfers, is responsible for the
  cancellation of singularity.}
\end{figure}

From our analysis follows that small momentum transfers $q_{1,2}$ are important for the
appearance of the forward peak. In this regime the Coulomb interaction is screened by
plasmons suggesting that the inclusion of screening may reduce the peak. Therefore, we
performed three calculations for $k=0.85\kF$ with i) bare Coulomb lines, ii) with only
plasmon screening, and iii) using the fully screened RPA $W_0$. They indeed demonstrate
that the plasmonic contribution to $\Sigma_{a\bar{a}}$ is essential for compensating the
singularity in the bare Coulomb term (Fig.~\ref{fig:SGMa_v_pl_tot}). Notice that they both
appear with the same sign because the interactions enter quadratically in the expression
for $\Sigma_{a\bar{a}}$. As a result, a smooth frequency dependence free of any
singularities is obtained for the sum of all contributions.

\subsection{Cancellations between $\Sigma_{aa}^<$ and $\Sigma_{a\bar{a}}^<$ in the
  asymptotic regime\label{sec:asympt}} The asymptotic regime $\omega\rightarrow\pm\infty$
is important because there $\Sigma_{aa}$ approaches zero, and a failure of the PSD
construction would be evident.  It is convenient to perform derivations using scaled
variables $\vec{x}_{i}=\vec{k}_{i}/\kF$, $\vec{y}_{i}=\vec{q}_{i}/\kF$, $\zeta=\w/\eF$.
\paragraph{3D HEG}
The $G_0W_0$ self-energy scales in the high-frequency limit~\cite{pavlyukh_time_2013} as
\begin{align}
  \tfrac{\ii}{2}\overline{\Sigma}_{aa}^{>}(x,\zeta)&\equiv
  \tfrac{\ii}{2}\Sigma_{aa}^{>}(k/\kF,\w/\eF)/\eF
  \stackrel{\zeta\rightarrow\infty}{\longrightarrow}
  c_1\frac{\alpha_3^2r_s^2}{\zeta^{3/2}},\\
  c_1&=\frac{16\sqrt{2}}{3\pi}.
\end{align}
Conversely, this determines the short-time behavior of the electron GF. Unexpectedly, Vogt
\emph{et al.}~\cite{vogt_spectral_2004} have demonstrated that the second-order exchange
$\Sigma_{2x}$ asymptotically scales in the same way, but with an additional $-\nicefrac12$
prefactor. This result can be further generalized and derived as follows.

\begin{figure}[t] 
  \centering
\includegraphics[width=\columnwidth]{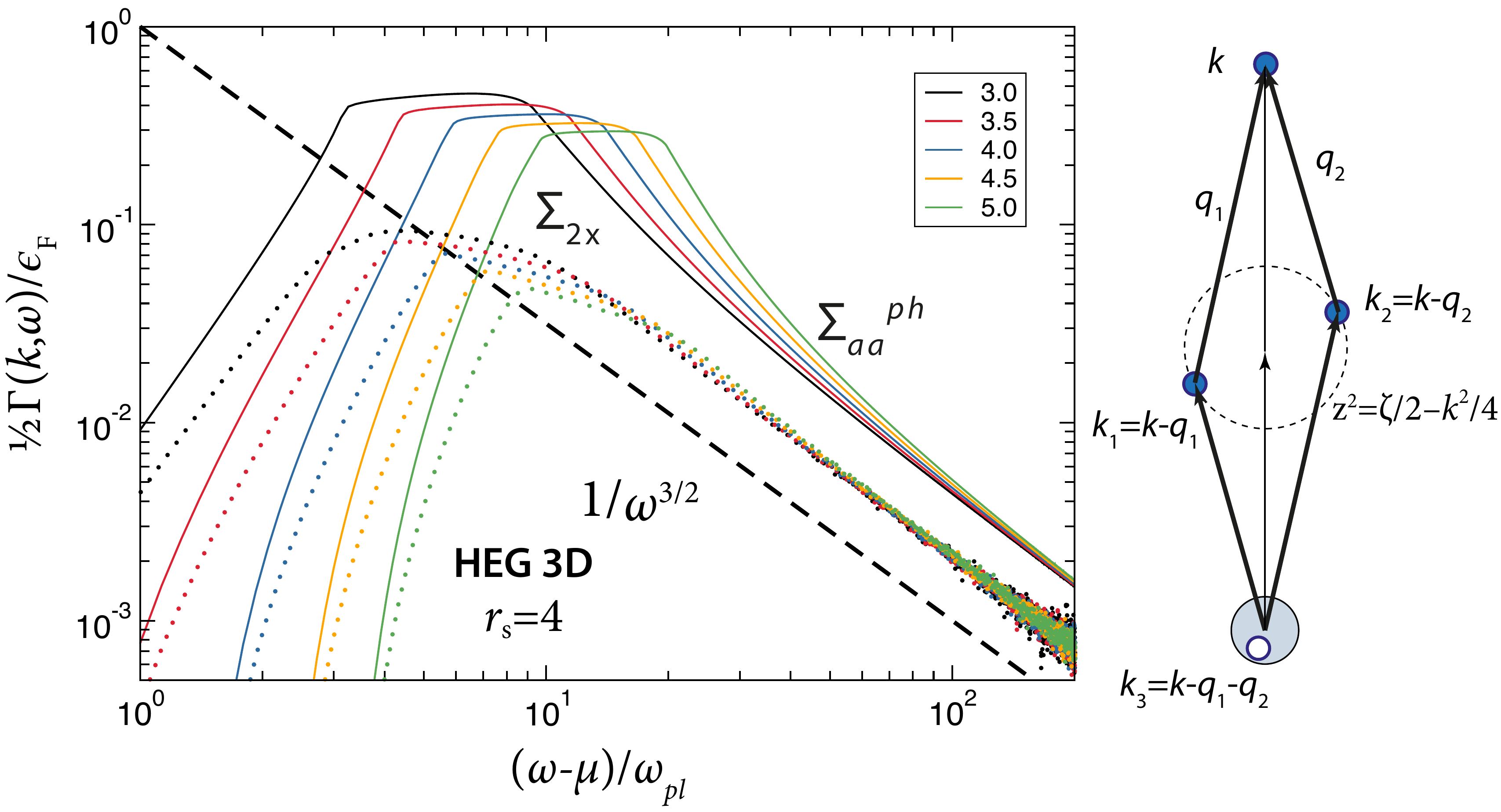}
\caption{\label{fig:asymp} Asymptotic behavior of $\ii \Sigma_{aa}^>$ (only $p$-$h$
  excitations) and $-\ii \Sigma_{2x}^>$ for large momentum values plotted on the logarithmic
  scale. The latter quantity is negative and therefore is multiplied with $-1$. A scheme
  on the right illustrates momentum configuration of $p\rightarrow p+p+h$ scattering at
  large energy and momentum transfer, i.\,e. in the asymptotic regime $\w\gg
  k^2/2\gg\eF$. In this limit $k_{3}\ll k$ and can be integrated over.}
\end{figure}

The generalization concerns the fact that in the high frequency limit, the screening is
not important and the screened interactions in the expression for $\Sigma_{a\bar{a}}$ can
be replaced with the bare Coulomb, i.\,e., $W^{--}(q_{1},\w-\epsilon_1)\rightarrow
v(q_{1})$, $W^{++}(q_{2},\w-\epsilon_2)\rightarrow v(q_{2})$. This means that
$\Sigma_{a\bar{a}}$ asymptitocally behaves as $\Sigma_{2x}$. Using the momenta flow as in
Fig.\,\ref{fig:diag2}(a), the second-order exchange reads
\begin{align}
  \tfrac{\ii}{2}\overline{\Sigma}_{2x}^{>}(x,\zeta)&
  =-\frac{\alpha_3^2 r_s^2 }{\pi^3}\iint \frac{\dd^3 \vec{y}_{1}}{y_{1}^2} \frac{\dd^3 \vec{y}_{2}}{y_{2}^2}\,
  \nbF(x_{1})\nbF(x_{2})\nF(x_{3})\nn\\
  &\qquad\qquad \times \delta(\zeta-\epsilon_1-\epsilon_2+\epsilon_3).
\end{align}
In the asymptotic case $\zeta\gg x^2/2\gg1$ we have $x_{3}\ll x_{1,2}$.  We change
the variables
\begin{align*}
\vec{y}_{1,2}=\frac12\vec{x}\pm\vec{z}=\vec{x}_{2,1}
\end{align*}
and integrate over $\vec{x}_{3}$ within the Fermi sphere (from the scheme in
Fig.~\ref{fig:asymp} it is evident that to a good approximation the integrand is
independent of $x_{3}$) yielding the $4\pi/3$ prefactor to the following remaining
integral
\begin{align}
  \tfrac{\ii}{2}\overline{\Sigma}_{2x}^{>}(x,\zeta)&\approx-\frac{4\alpha_3^2 r_s^2}{3\pi^2}\int\!
  \frac{\dd^3\vec{z}}{\big|\frac12\vec{x}-\vec{z}\big|^2\big|\frac12\vec{x}+\vec{z}\big|^2}
  \delta\big(\zeta-\frac12x^2-2z^2\big)\nn\\
  &=-\frac{8\alpha_3 ^2 r_s^2}{3\pi}\frac{ \tanh ^{-1}\mleft(\frac{x \sqrt{2 \zeta -x^2}}{\zeta
    }\mright)}{\zeta x} \nn\\
  &=-\frac{8\sqrt{2}}{3\pi}\frac{\alpha_3 ^2 r_s^2}{\zeta^{3/2}}+\mathcal{O}(x^2).
  \label{eq:est}
\end{align}
In the last step, we exploit the high-frequency assumption $\zeta\gg x^2/2$ and perform
a series expansion over $x$ to get the conjectured scaling
$\ii \overline{\Sigma}_{2x}^{>}(x,\zeta)\stackrel{\zeta\rightarrow\infty}{\longrightarrow}
c_2\alpha_3^2r_s^2 \zeta^{-3/2}$.  The scaling is verified numerically in
Fig.~\ref{fig:asymp} confirming that the constant $c_2$ computed from Eq.~\eqref{eq:est}
is momentum-independent.  It is important, however, that $c_2=-\frac12c_1$ ensuring the
PSD property in the high-frequency limit.

\paragraph{2D HEG}
\begin{figure}[] 
  \centering
\includegraphics[width=\columnwidth]{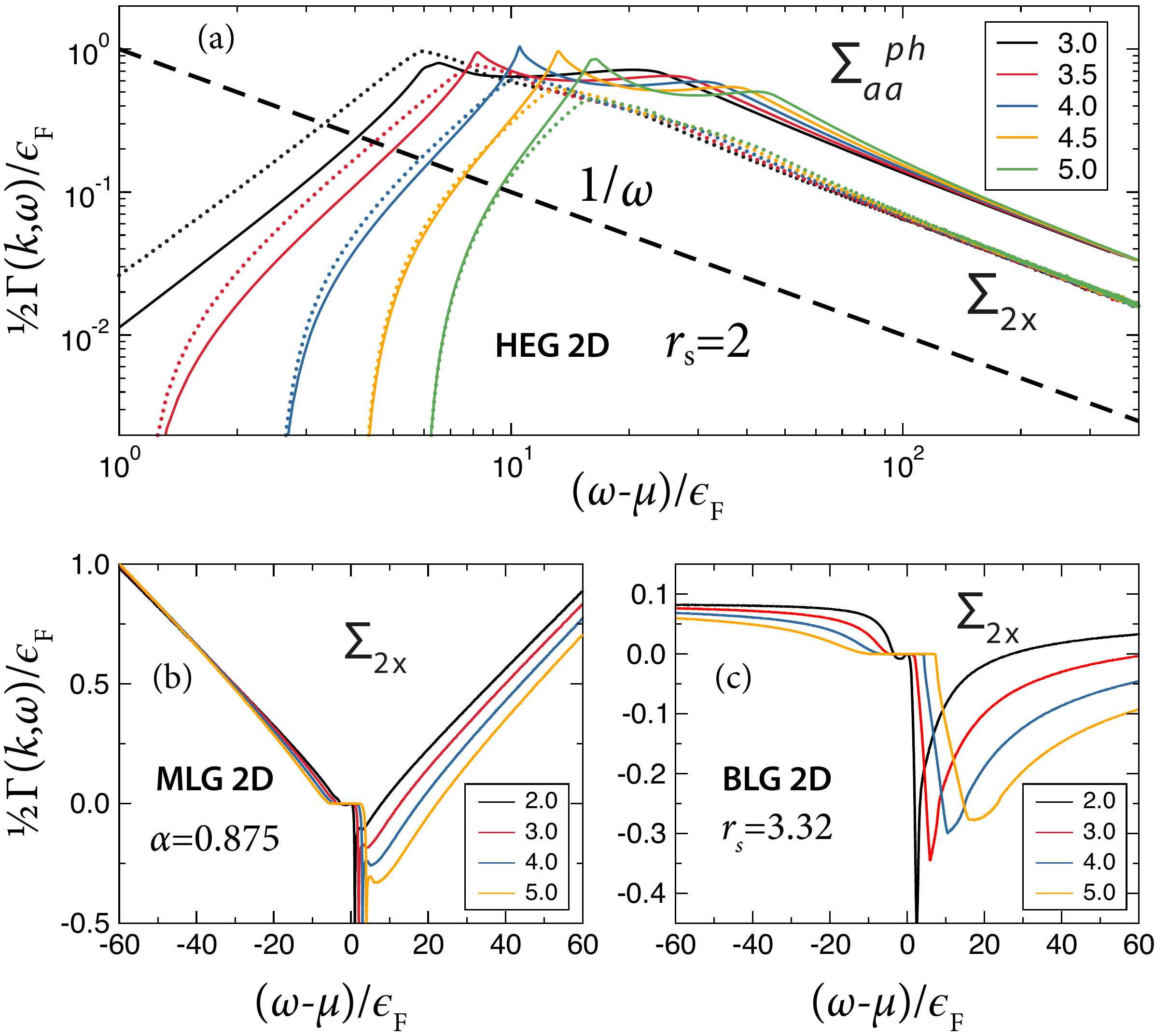}
\caption{\label{fig:asymp:2D} Top: asymptotic behavior of the rate functions
  $\ii \Sigma_{aa}^>$ (only $p$-$h$ excitations) and $-\ii \Sigma_{2x}^>$ for large momentum
  values plotted on the logarithmic scale for 2D HEG. As in the 3D HEG, they posses the
  $c_{1,2}/\w$ scaling, respectively, with $c_2=-1/2c_1$. In panels (b,c) the rate
  function of the two graphene systems is plotted for different momentum values. There is
  no universal scaling.}
\end{figure}
The derivation follows the same line:
\begin{align}
  \tfrac{\ii}{2}\overline{\Sigma}_{2x}^{>}(x,\zeta)&\approx -2\int\!
  \frac{\dd^2\vec{z}}{\big|\frac12\vec{x}-\vec{z}\big|\big|\frac12\vec{x}+\vec{z}\big|}
  \delta\big(\zeta-\frac12x^2-2z^2\big)\nn\\
  &=-\frac{4}{|\zeta-x^2|}K\mleft(\frac{x\sqrt{x^2-2\zeta}}{\zeta-x^2}\mright)\nn\\
  &=-\frac{2\pi}{\zeta}+\mathcal{O}(x^2),
\end{align}
where $K$ is a complete elliptic integral. As in the case of the 3D HEG, there is a
universal (momentum-independent) asymptotic scaling, see Fig.~\ref{fig:asymp:2D}(a), and
the ratio of the prefactors is the same. This is yet another exact analytical statement
about the second-order self-energy.
\paragraph{Graphene systems}
\begin{figure}[] 
  \centering
\includegraphics[width=\columnwidth]{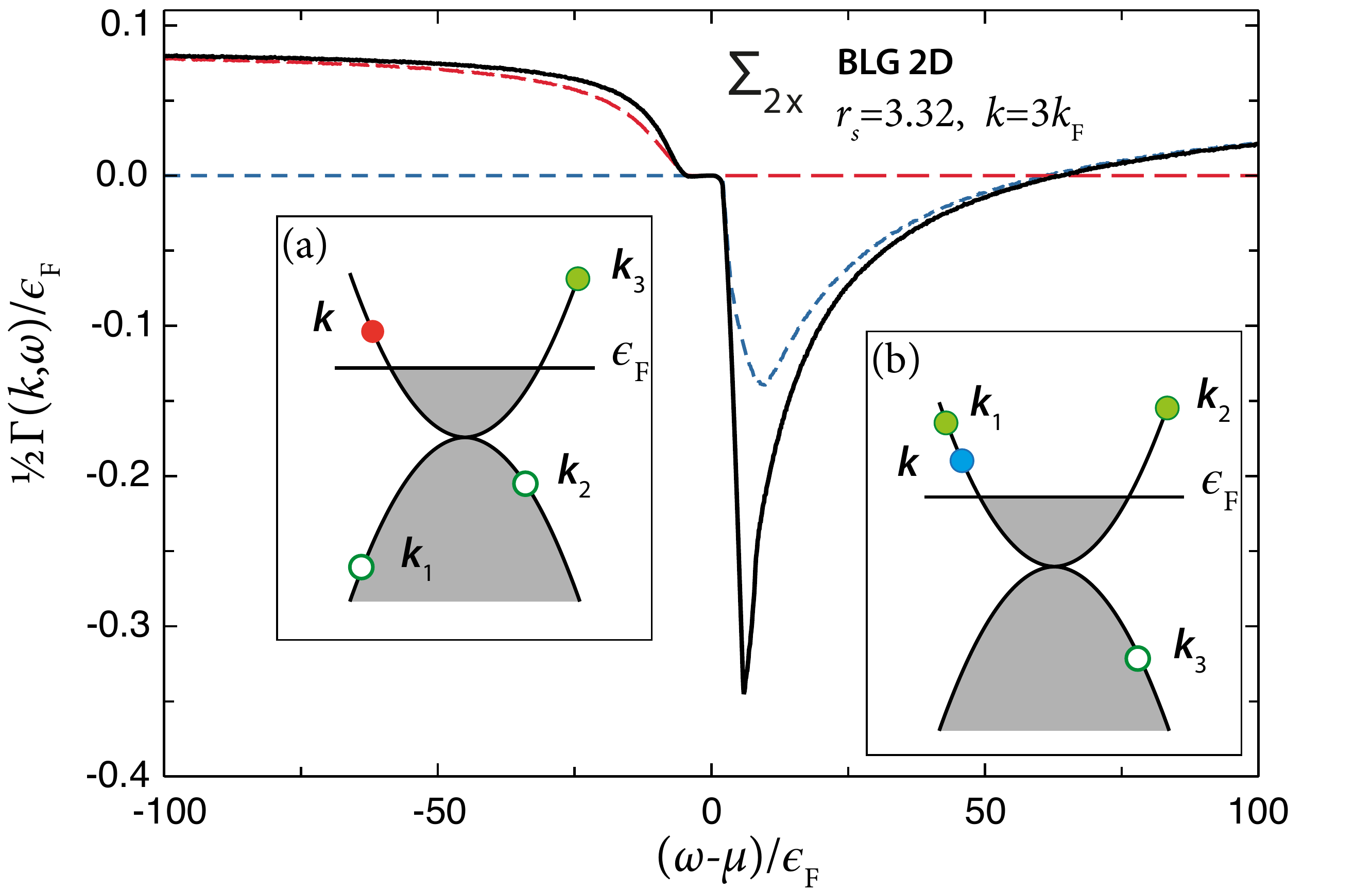}
\caption{\label{fig:S2x:BLG} Second-order exchange in the BLG system for $k=3\kF$. Besides
  common scattering channels, the $\w\rightarrow\infty$ behavior is dominated by a
  processes (blue line) with one hole in the lower band, $s_3=-1$ (inset b), for
  $\omega\rightarrow-\infty$ a mechanism with 2 holes in the lower band $s_1=s_2=-1$ is
  dominating (inset a).}
\end{figure}
The situation is much more complex in the case of graphene,
Fig.~\ref{fig:asymp:2D}(b,c). Besides the usual scattering processes considered above,
$\Sigma_{a\bar{a}}$ contains processes in which particles change the band,
Fig.~\ref{fig:S2x:BLG}. For instance, for $k>\kF$ our calculations indicate that
$\Sigma_{2x}(k,\w)$ is dominated by the process with $s_3=-s_1=-s_2=1$ for $\w<0$, and
with $-s_3=s_1=s_2=1$ for $\w>0$, see Eq.~\eqref{eq:F:2x}. This leads to the scattering
rates that do not tend to zero as $\w\rightarrow \pm\infty$. They are cut-off dependent
and should be treated as in the case of the first-order exchange.
\subsection{Quasiparticle properties\label{sec:qp}}
\begin{figure}[] 
  \centering
\includegraphics[width=\columnwidth]{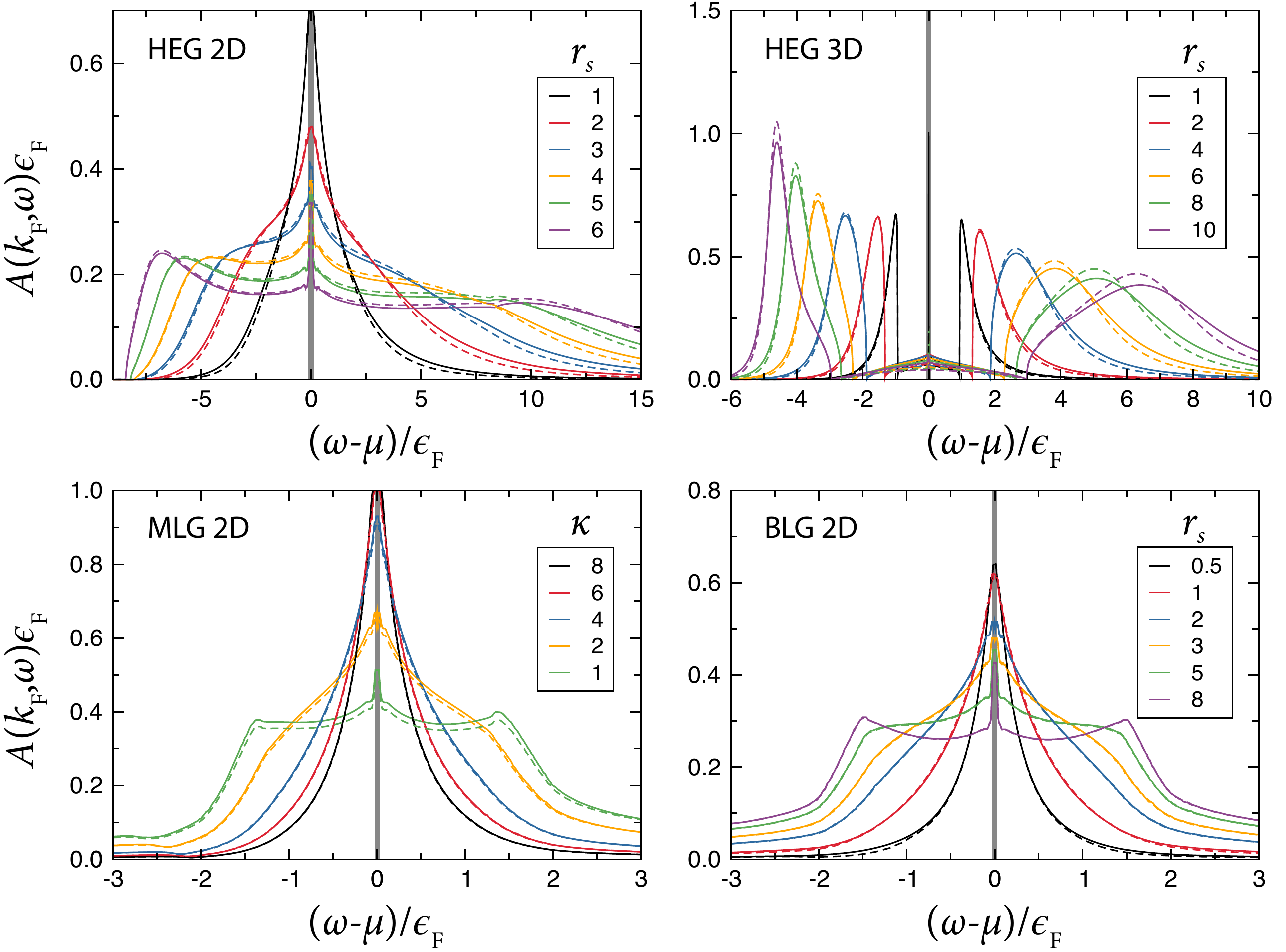}
\caption{\label{fig:AkF} Electron spectral functions for quasiparticle states at the Fermi
  surface ($k=\kF$) for different interaction strengths. Dashed lines stand for the
  $G_0W_0$ results, full lines\,---\,additionally include $\Sigma_{a\bar{a}}$.}
\end{figure}
All four considered systems possess very distinct spectral functions.  In the vicinity of
the quasiparticle peak they can be represented  in the Lorentzian form
\begin{align}
  A(k,\w)&=Z_{qp}\frac{1/\tau(k)}{(\w-\epsilon_{qp}(k))^2+1/(2\tau(k))^2},
  \label{eq:A:qp}
\end{align}
where the peak position $\epsilon_{qp}(k)$, the inverse life-time $1/\tau(k)$, and the
quasiparticle renormalization factor $Z_{qp}$ can be determined by solving the Dyson
equation with the given retarded self-energy operator (see Sec.~13.1 in
Ref.~\cite{stefanucci_nonequilibrium_2013}). In general, one has to take care whether the
spectral density can really be written in this form with \emph{finite} $Z_{qp}$. For
instance, this might be not the case in undoped
graphene~\cite{gonzalez_marginal-fermi-liquid_1999,barnes_effective_2014}, or 2D systems
with short-range repulsive interactions~\cite{chubukov_kohn-luttinger_1993}, but \emph{is}
the case for the systems considered here.  As can be seen from Fig.~\ref{fig:AkF}, the
second-order self-energy has a rather small impact on the shape of quasiparticle peak and
its satellites. Therefore, in order to quantify the effect we compute the quasiparticle
peak strength:
\begin{equation}
  Z_{qp}(k)=\left(1-\frac{\partial}{\partial \w}
  \left.\re\SgR(k,\w)\right|_{\w=\epsilon_{qp}(k)}\right)^{-1}.
\end{equation}
We further characterize the quasiparticle dispersion in terms of the effective mass:
\begin{equation}
  \frac1{m^{*}}=\frac1{\kF}\left.\frac{\dd \epsilon_{qp}(k)}{\dd k}\right|_{k=\kF},
  \label{eq:m:eff:def}
\end{equation}
and the Fermi velocity (for MLG)
\begin{equation}
  \vF^{*}=\left.\frac{\dd \epsilon_{qp}(k)}{\dd k}\right|_{k=\kF}.
\end{equation}
Finally, the inverse quasiparticle life-time is computed,
\begin{align}
  \tau(k)^{-1}&=Z_{qp}(k)\Gamma(k,\epsilon_{qp}(k)),\label{eq:tau:def}\\
  \tfrac{1}{2}\Gamma(k,\w)&=-\im\SgR(k,\w). 
\end{align}

We are mostly interested in the correlated regime $r_s\gg1$. However, the asymptotic
results $r_s\rightarrow0$ are also shown when available in order to demonstrate that they
are valid in a rather very narrow density interval.  Our main comparison is with three
classes of theories. As benchmarks for the homogeneous electron gas, the quantum
Monte-Carlo results of Holzmann \emph{et al.}~\cite{holzmann_renormalization_2009} (2D)
and \cite{holzmann_momentum_2011} (3D) are used. The second class of methods has been
advocated by Giuliani and co-workers: Ref.~\cite{asgari_quasiparticle_2005} (2D) and
Ref.~\cite{simion_many-body_2008} (3D). They improve upon $G_0W_0$ by using parameterized
data from QMC calculations in terms of the charge and the spin static local fields
factors, $G_+(q)$ and $G_-(q)$, respectively. In their method, the self-energy is in the
$GW$ form, however, the screened interaction is replaced by the Kukkonen-Overhauser
effective interaction~\cite{kukkonen_electron-electron_1979}. Furthermore, a diagrammatic
approach based on the Bethe-Salpeter equation for the improved screened interaction by
Kutepov and Kotliar~\cite{kutepov_one-electron_2017} is also used for
comparison. Unfortunately, none of these theories are available for graphene systems.
\begin{figure}[] 
  \centering
\includegraphics[width=\columnwidth]{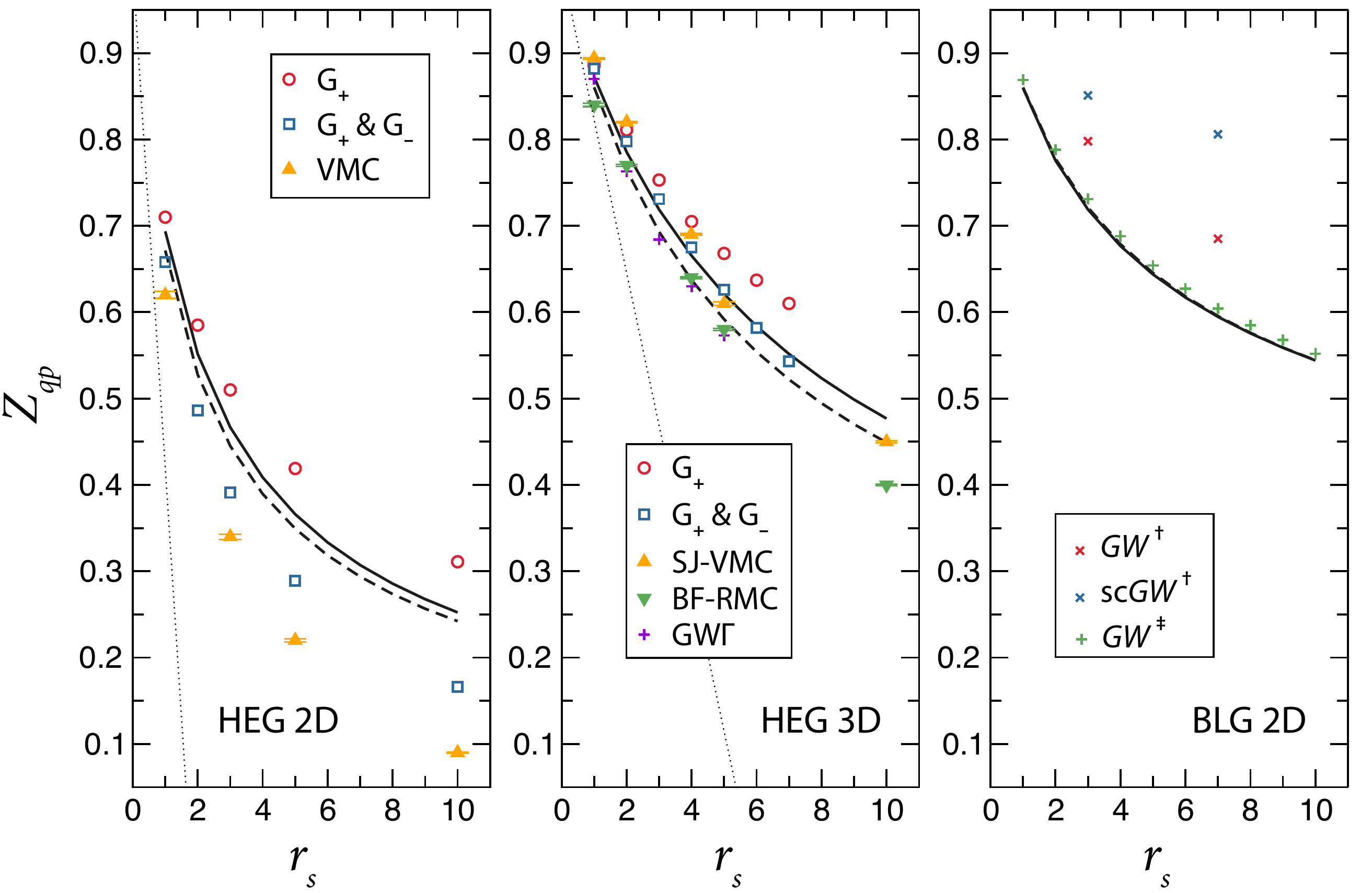}
\caption{\label{fig:Zqp} Quasiparticle renormalization factor for $k=\kF$ and different
  interaction strengths: $\Sigma_{G_0W_0}$---dashed,
  $\Sigma_{G_0W_0}+\Sigma_{a\bar{a}}$---solid lines.  For 2D HEG, the variational
  Monte-Carlo (VMC) results are taken from Ref.~\cite{holzmann_renormalization_2009}, and
  the VMC calculations using Slater-Jastrow (SJ) wave-functions and backflow (BF)
  reptation Monte-Carlo (RMC) for 3D HEG from Ref.~\cite{holzmann_momentum_2011}. $G_+(q)$
  and $G_+(q)\& G_-(q)$ denote methods based on the local structure factors from
  Ref.~\cite{asgari_quasiparticle_2005} (2D) and Ref.~\cite{simion_many-body_2008}
  (3D). $GW\Gamma$ is a scheme D of Ref.~\cite{kutepov_one-electron_2017}, it includes the
  first-order vertex function in the finite temperature formalism. For BLG,
  $\dagger$\,---\,Ref.~\cite{sabashvili_bilayer_2013}, $\ddagger$\,---\,
  Ref.~\cite{sensarma_erratum:_2012}.}
\end{figure}

We start by compiling the data for homogeneous electron gases in 2D and 3D and the bilayer
graphene. In these systems $r_s$ is the relevant parameter that can be controlled by
doping or other means. In MLG, there is only an indirect possibility to control $\alpha$
by changing the background dielectric constant and $\kappa$. This system will be considered
later.
\paragraph{$r_s$ as a control parameter}
In Fig.~\ref{fig:Zqp} the quasiparticle renormalization factor $Z_{qp}(\kF)$ as a function
of the density parameter $r_s$ is shown. As expected, the agreement between different
methods deteriorates with increasing $r_s$, moreover the quantum Monte-Carlo results are
not available for BLG.  Therefore, it is hard to say with absolute certainty what is the
``right'' value. On the positive side, we see a very nice convergence of all methods
towards the linear asymptote
\begin{align}
  Z_{qp}&=1-\left(\frac12+\frac1\pi\right)\alpha_2 r_s,& 2D\\
  Z_{qp}&=1-\frac{c}{\pi^2}\alpha_3r_s,& 3D.
\end{align}
with $c=-\int_0^{\pi/2}\log(1-x\cot x)\,\dd x\approx3.353$.  The 3D result is by Daniel
and Vosko~\cite{daniel_momentum_1960}, and the 2D asymptote together with temperature
corrections is due to Galitski and Das Sarma~\cite{galitski_universal_2004}. For BLG, the
effect of $\Sigma_{a\bar{a}}$ is negligible, and our calculations accurately reproduce the
corrected results of Sensarma \emph{et al.}~\cite{sensarma_erratum:_2012}, whereas the
one-shot and the self-consistent $GW$ calculations of Sabashvili \emph{et
  al}.~\cite{sabashvili_bilayer_2013} deviate. For HEG in 2D and 3D, the inclusion of
$\Sigma_{a\bar{a}}$ slightly increases the value of $Z_{qp}$. It is interesting to notice
that the same trend is observed when the charge local field factor $G_+(q)$ is
included. For HEG 2D, the additional inclusion of both local fields reduces $Z_{qp}$ in
agreement with variational MC calculations. At variance, for HEG 3D the effect of the spin
local field is less pronounced~\cite{simion_many-body_2008}, and is nearly the same as in
our calculations with $\Sigma_{a\bar{a}}$. The effect of the vertex function in
Ref.~\cite{kutepov_one-electron_2017} is rather small, therefore, it would be interesting
if these calculations could be extended towards larger $r_s$, where even variational and
backflow reptation MC results are in disagreement.

\begin{figure}[] 
  \centering
\includegraphics[width=\columnwidth]{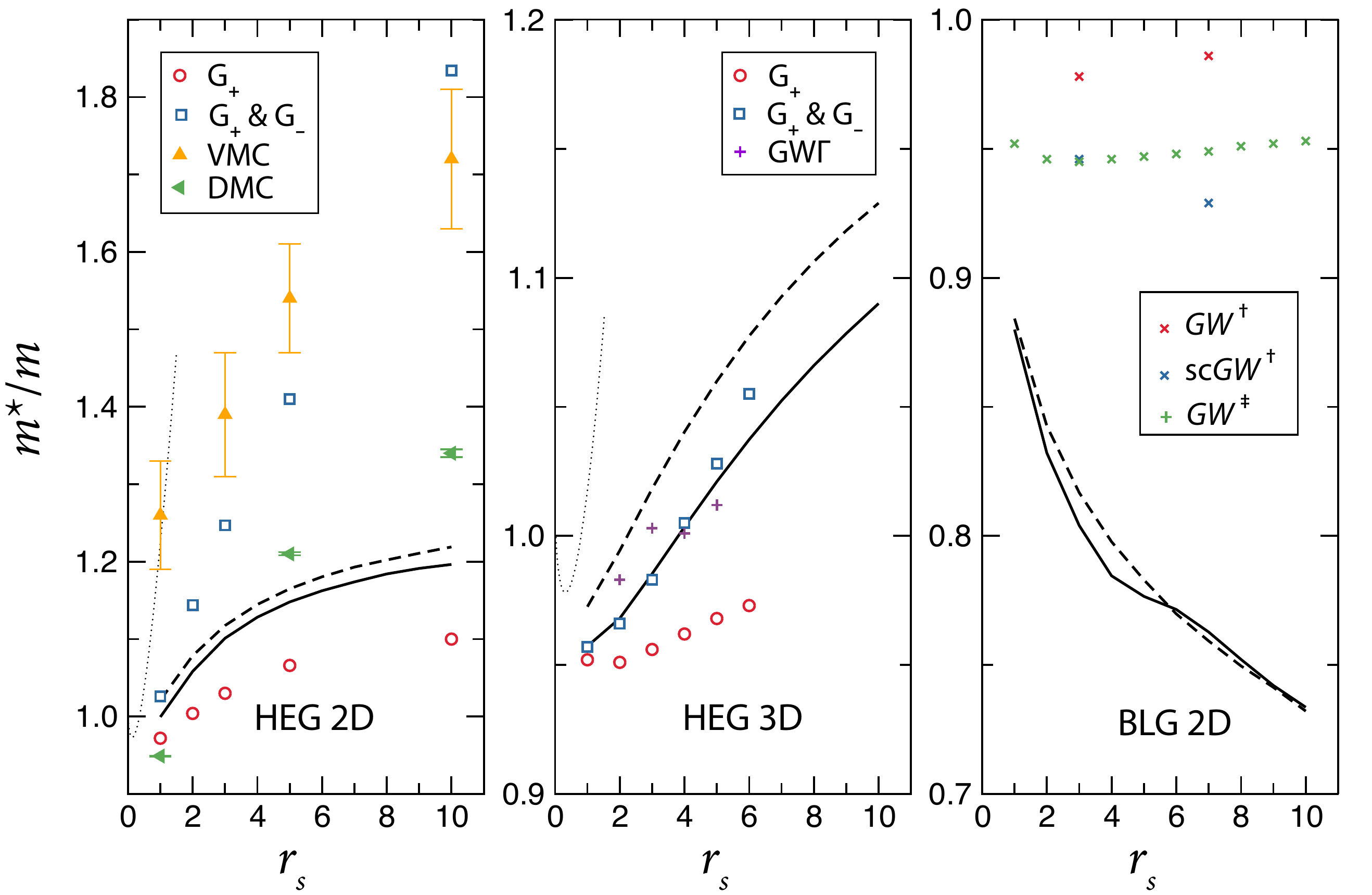}
\caption{\label{fig:m} Effective mass for different interaction strengths:
  $\Sigma_{G_0W_0}$---dashed, $\Sigma_{G_0W_0}+\Sigma_{a\bar{a}}$---solid lines.  For 2D
  HEG, the variational Monte-Carlo (VMC) results are taken from
  Ref.~\cite{holzmann_renormalization_2009}, and the DMC calculations from
  Ref.~\cite{drummond_quantum_2009}. $G_+(q)$ and $G_+(q)\& G_-(q)$ denote methods based
  on the local structure factors from Ref.~\cite{asgari_quasiparticle_2005} (2D) and
  Ref.~\cite{simion_many-body_2008} (3D). $GW\Gamma$ is a scheme D of
  Ref.~\cite{kutepov_one-electron_2017}, it includes the first-order vertex function in
  the finite temperature formalism. For BLG,
  $\dagger$\,---\,Ref.~\cite{sabashvili_bilayer_2013}, $\ddagger$\,---\,
  Ref.~\cite{sensarma_erratum:_2012}.}
\end{figure}

Less accurate are the predictions of different theories for the effective mass,
Fig.~\ref{fig:m}. The situation gets complicated due to different methods of its
determination adopted in
literature~\cite{zhang_quasiparticle_2005,asgari_quasiparticle_2005}. In order to avoid
any ambiguities, the masses in our approach are obtained per definition, that is by
solving the Dyson equation for $\epsilon_{qp}(k)$ and using Eq.~\eqref{eq:m:eff:def}, and
\emph{not} by using the self-energy representation
\begin{equation}
  \frac1{m^{*}}=\frac{Z_{qp}^{-1}}{1+\frac{1}{\kF}
    \left.\frac{\dd \re\SgR(k,\eF)}{\dd k}\right|_{k=\kF}}.
  \label{eq:m:eff:def2}
\end{equation}

For weakly interacting systems, $r_s\rightarrow0$ (high-density limit), we compare with
asymptotic expansions. A general form~\cite{zhang_quasiparticle_2005} valid for 2D and 3D
homogeneous electron gases reads
\begin{align}
m^{*}&=1+a r_s(b+\log r_s).
\end{align}  
The coefficients $a$ and $b$ in three dimensions can be inferred from the well-known
result of Gell-Mann~\cite{gell-mann_specific_1957} for the specific heat.  The correction
in the linear temperature-dependent term due to the electron-electron interaction is
entirely attributed to the mass renormalization~\cite{mahan_many-particle_2000}, and
therefore
\begin{align}
  m^{*}&=1+\frac12\frac{\alpha_3 r_s}{\pi}\left(\log\frac{\alpha_3 r_s}{\pi}+2\right).
\end{align}
In two dimensions, the original derivation is due to Janak~\cite{janak_g_1969}, whereas
the corrected formula\,\footnote{There has been some controversies in this derivation. For
  instance, some mistakes in the original result were pointed out
  Refs.~\cite{ting_effective_1975, ando_electronic_1982}, but not explicitly corrected;
  wrong coefficients $a$ and $b$ can be seen in
  Refs.~\cite{galitski_universal_2004,zhang_quasiparticle_2005}.} can be found in Saraga
and Loss~\cite{saraga_fermi_2005}
\begin{align}
  m^{*}&=1+\frac{\alpha_2 r_s}{\pi}\left(\log\frac{\alpha_2 r_s}{2}+2\right).
\end{align}
The asymptotic expressions are derived with the help of additional approximations
(e.\,g. static screening), which quickly invalidates them as $r_s$ increases, see dotted
lines in Fig.~\ref{fig:m}.

Let us now inspect the influence of $\Sigma_{a\bar{a}}$ on the effective mass, that is the
difference between the full and the dashed lines. One impressive observation is that for
3D HEG, our calculations agree again very well with results of Simion and
Giuliani~\cite{simion_many-body_2008}, where both local field factors are taken into
account. The charge local field alone tends to underestimate the effective mass for both
systems. A rather poor performance of the Monte-Carlo methods is evident for the 2D HEG as
well, further calculations of effective masses and extensive comparisons can be found in
Drummond and Needs~\cite{drummond_quantum_2009}. However, the difficulties to extract
excited state properties from these methods are understandable, and the work of Eich,
Holzmann and Vignale~\cite{eich_effective_2017} provides some justification.

For BLG, the mass renormalization substantially deviate in comparison with Sensarma
\emph{et al.}~\cite{sensarma_erratum:_2012}. This might be due to a different procedure
based on Eq.~\eqref{eq:m:eff:def2} adopted in this work. Large and negative mass
renormalization indicates that for $r_s\gg1$ the system goes into a correlated regime.

\paragraph{$\kappa$ as a control parameter}
\begin{figure}[] 
  \centering \includegraphics[width=\columnwidth]{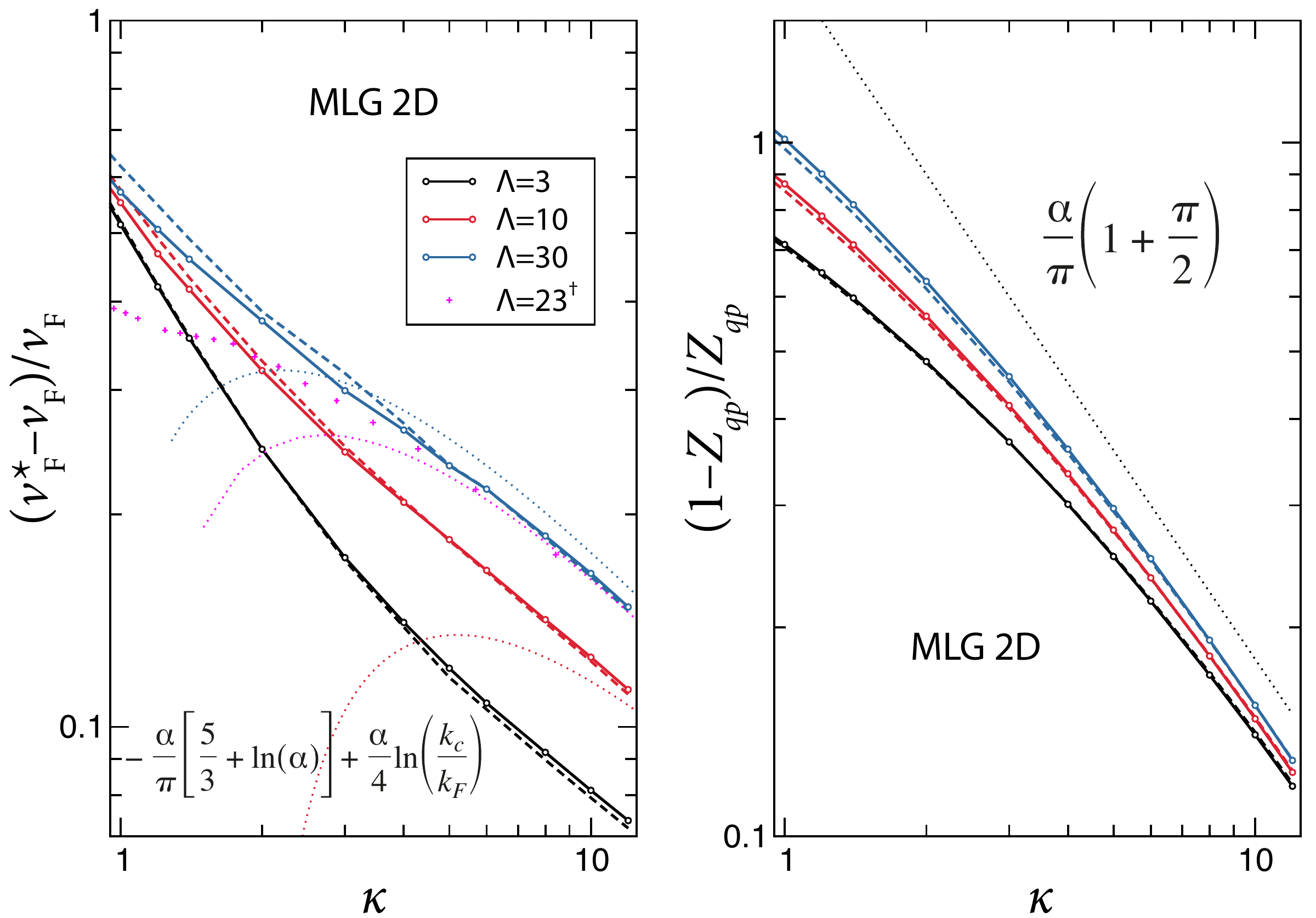}
\caption{\label{fig:vZ} Renormalized Fermi velocity and the quasiparticle renormalization
  factor of the monolayer graphene for different values of the momentum cut-off
  $\Lambda=k_c/\kF$ as a function of dielectric constant. The log-log plots are compared
  with the scalings from Ref.~\cite{das_sarma_many-body_2007} (dotted
  lines). $^\dagger$\,---\,data from Ref.~\cite{das_sarma_velocity_2013}.}
\end{figure}
Let us recapitulate first that we focus on the \emph{extrisic} monolayer graphene system
here, that is $\kF>0$. This is essentially a classical Fermi liquid in the marked contrast
to the more complicated \emph{intrinsic} graphene, $\kF=0$. For the latter, we refer to a
comprehensive summary by Tang \emph{et al.}~\cite{tang_role_2018}. While many conceptual
problems do not arise in the extrinsic case, some important insight can be obtained from
the intrinsic graphene.  Consider for instance the expression for the Fermi velocity
renormalization derived in \cite{das_sarma_many-body_2007} and plotted in
Fig.~\ref{fig:vZ} (left) as dotted lines 
\begin{align}
  \frac{\vF^{*}-\vF}{\vF}&=-\frac{\alpha}{\pi}\left(\frac53+\log{\alpha}\right)
  +\frac{\alpha}4\log\mleft(\frac{k_c}{\kF}\mright).
  \label{eq:veff:asymp}
\end{align} 
Here, the first part is extrinsic. It describes scattering processes in which the initial
and the final state belong to the same band $s=s'=1$ and contains no adjustable
parameters. The second part is intrinsic, it includes scattering processes changing the
band $s=-s'=1$ and therefore depends on the momentum cut-off. In going to higher
perturbative orders, such as including $\Sigma_{a\bar{a}}$, more and more processes
involve interband scatterings and the role of intraband scattering is diminishing (as
explicitly demonstrated for BLG, Fig.~\ref{fig:S2x:BLG}).

Let us inspect the qualitative dependence of the renormalized velocity on the interaction
parameter $\alpha\simeq2.2/\kappa$~\eqref{eq:MLG:alpha:val}. At higher $\alpha$, the
dependence deviates from linear. This is already evident from the extrinsic part in
Eq.~\eqref{eq:veff:asymp}. The intrinsic part shows a similar trend when computed beyond
the leading order~\cite{tang_role_2018}. By consistently including other terms, one can
improve the agreement of asymptotic theory with our numerical results.

Generally, it is believed that $G_0W_0$ result are already very
accurate~\cite{das_sarma_velocity_2013}. However, higher-order diagrams have been treated
in Ref.~\cite{barnes_effective_2014}, quantum Monte-Carlo calculations were performed in
Ref.~\cite{tang_role_2018}. All of them concern with intrinsic case, which is still
relevant to some extent as stated above, however cannot be used for a direct
comparison. They predict a slightly larger velocity renormalization, whereas, we observe
here that the inclusion of $\Sigma_{a\bar{a}}$ leads to smaller values, Fig.~\ref{fig:vZ}.
\paragraph{Quasiparticle life-time}
\begin{figure}[] 
  \centering \includegraphics[width=\columnwidth]{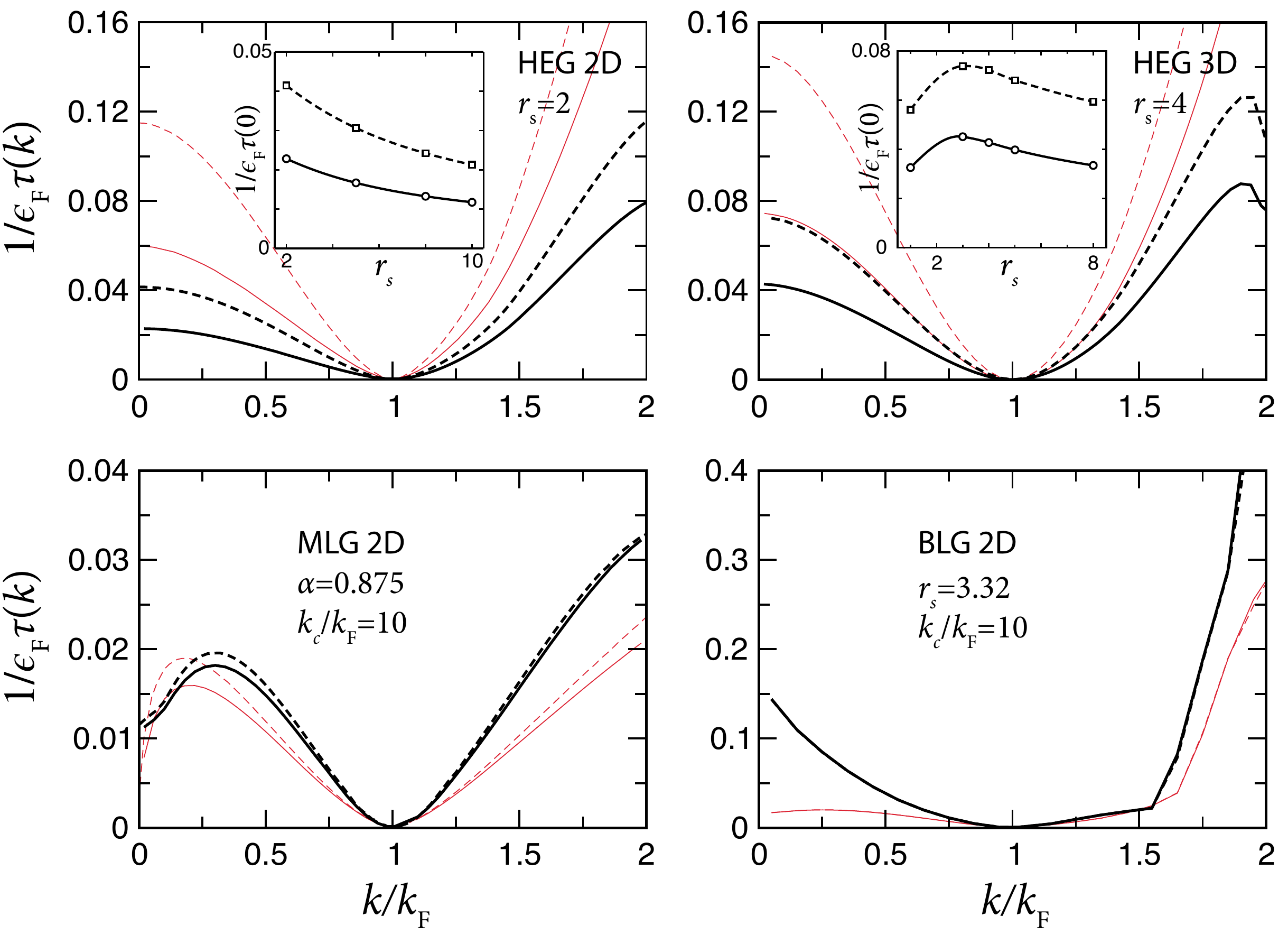}
\caption{\label{fig:Gm} Inverse quasiparticle life-time~\eqref{eq:tau:def} as a function
  of momentum. In the two insets, the values at the band bottom $\tau(k=0)^{-1}$ is shown
  for different $r_s$. Red lines additionally show the on-shell
  approximation~\eqref{tau:approx} without (dashed) and with (full) $\Sigma_{a\bar{a}}$.}
\end{figure}
$1/\tau(k)$ is an essential ingredient of the quasiparticle spectral
function~\eqref{eq:A:qp}. We determine it by solving the Dyson equation in the complex
frequency plane. Also, it can be obtained from the Fermi Golden rule as advocated by Qian
and Vignale~\cite{qian_lifetime_2005}.  From the discussion in Sec.~\ref{sec:psd} we know
that the two are completely equivalent.

In Fig.~\ref{fig:Gm} we summarize our finding for $\tau(k)$ computed for $0<k<2\kF$. In
many studies, the ``on-shell'' imaginary self-energy is taken as a measure of the inverse
life-time
\begin{align}
  \tau(k)^{-1}&=-\im\SgR(k,\epsilon(k)),
  \label{tau:approx}
\end{align}  
where $\epsilon(k)$ is the bare dispersion relation. Apart from missing the $Z_{qp}$
prefactor, this approach is reasonable for $k\approx \kF$, where the difference between
the ``true'' $\epsilon_{qp}(k)$ and the bare $\epsilon(k)$ spectrum is small. For
$\xi_k=|\epsilon(k)-\eF|/\eF\gg1$ the difference between Eq.~\eqref{eq:tau:def} and the
on-shell approximation~\eqref{tau:approx} is substantial, as can be seen by comparing
black and red lines in Fig.~\ref{fig:Gm}. We find, for instance for MLG, that the
approximation incorrectly yields vanishing scattering rates at the Dirac point
($k=0$). One consequence of this is the diverging inelastic mean free path at zero
temperature predicted in Ref.~\cite{li_finite_2013}. On the other hand,
Eq.~\eqref{eq:tau:def} yields a finite value. It is worth noting that for the two HEG
systems the correction upon the on-shell value is mostly associated with the quasiparticle
strength $Z_{qp}$ renormalization, whereas, for graphene systems (MLG and BLG) the
deviation of $\epsilon_{qp}(k)$ from $\epsilon(k)$ also plays a role.

\begin{figure}[] 
  \centering \includegraphics[width=\columnwidth]{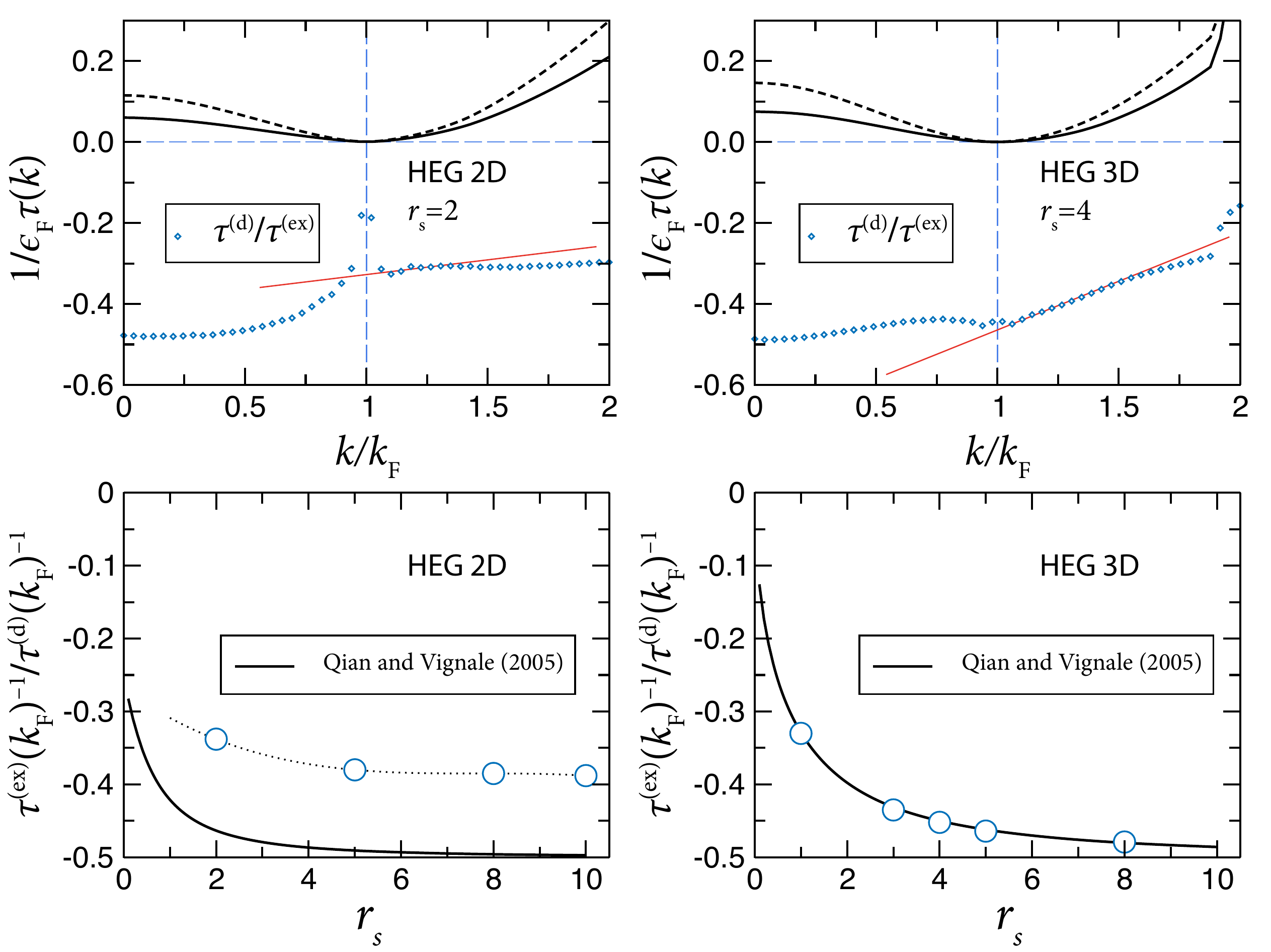}
\caption{\label{fig:Gm:ratio} Top row: determination of the ratio
  $\tau^\text{(ex)}(k)^{-1}/\tau^\text{(d)}(k)^{-1}$ for the homogeneous electron gas in
  2D and 3D illustrating numerical difficulties of taking the $k\rightarrow\kF$
  limit. Bottom row: comparison with analytical results of Ref.~\cite{qian_lifetime_2005}:
  $\zeta_n^\text{(ex)}(r_s)/\zeta_n^\text{(d)}(r_s)$, see
  Eq.~(\ref{eq:zeta:2d},\ref{eq:zeta:3d}).}
\end{figure}

Only for the two HEG systems the impact of $\Sigma_{a\bar{a}}$ is appreciable as depicted
in the insets of Fig.~\ref{fig:Gm} for $\tau(0)^{-1}$ for different $r_s$ values. The
dependence is not always monotonic. For $\xi_k\ll1$, we can compare with the asymptotic
expressions from Ref.~\cite{giuliani_quantum_2005}
\begin{align}
  \frac{1}{\tau(k)}&=\frac{1}{4\pi}\zeta_2(r_s)\xi_k^2\log\mleft(\frac{4}{\xi_k}\mright),&2D;\label{eq:tau:2d}\\
  \frac{1}{\tau(k)}&=\frac{\pi}{8}\zeta_3(r_s)\xi_k^2,&3D.\label{eq:tau:3d}  
\end{align}
They have shown that the inclusion of exchange modifies the density-dependent prefactors
$\zeta_n$ without affecting the functional form. However, the fitting of small values with
this form is not a trivial task because of (i) numerical issues and (ii) our insufficient
knowledge of the subleading terms\footnote{In 2D, there are disagreements in the
  subleading
  terms~\cite{zheng_coulomb_1996,reizer_electron-electron_1997,qian_lifetime_2005}.}.  We
follow Qian and Vignale~\cite{qian_lifetime_2005}, where the coefficients $\zeta_n(r_s)$
are derived:
\begin{align}
  \zeta_2(r_s)&=\left[1+\frac12\frac{\alpha_2^2r_s^2}{(1+\alpha_2 r_s)^2}\right]-
  \left[\frac14+\frac12\frac{\alpha_2r_s}{1+\alpha_2 r_s}\right],\label{eq:zeta:2d}\\
  \zeta_3(r_s)&=\frac{1}{2\lambda}\left[\tan^{-1}{\lambda}
    +\frac{\lambda}{1+\lambda^2}\right]\nn\\
  &-\frac{1}{2\lambda}\left[\frac{1}{\sqrt{2+\lambda^2}}
    \cot^{-1}\frac{1}{\lambda\sqrt{2+\lambda^2}}\right],\label{eq:zeta:3d}
\end{align}
with $\lambda=\pi^{1/2}/(\alpha_3 r_s)^{1/2}$. Here, the first brackets originate from the
direct (d) processes and the second\,---\,from the exchange (ex).  In
Fig.~\ref{fig:Gm:ratio}, we determine the ratio of exchange to direct scattering rates
using the on-shell approximation~\eqref{tau:approx} with $\Sigma_{a\bar{a}}$ and
$\Sigma_{G_{0}W_{0}}$, respectively. In 3D, the agreement with the analytical
results~\eqref{eq:tau:2d} is very good, whereas, in 2D we find that the ratio is
smaller. One possible explanation of this discrepancy could be the absence of the
plasmonic contributions to the direct scattering in Ref.~\cite{qian_lifetime_2005}.
Besides ratios, we also compared absolute values with analytical predictions and found
systematic underestimates. However, this should not be a surprise because the theory only
yields the leading terms.

\begin{figure}[] 
  \centering \includegraphics[width=\columnwidth]{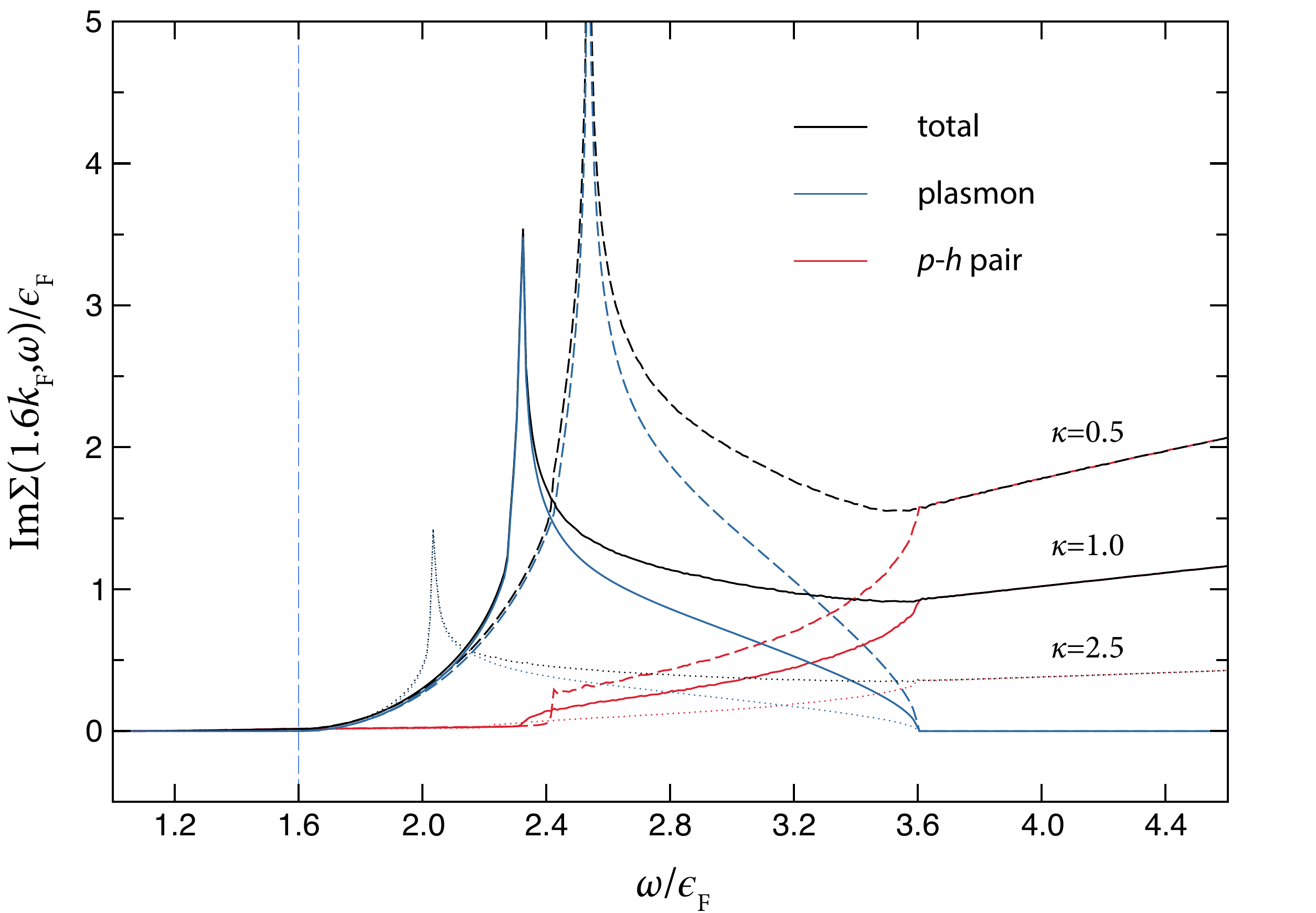}
\caption{\label{fig:Im:SGM:kp} Imaginary self-energy part of MLG for different interaction
  strengths. The on-shell value is only weakly dependent on $\kappa$ in agreement with the
  analytical result of Polini and Vignale~\cite{polini_quasiparticle_2014}.}
\end{figure}

For MLG, the inverse life-time follows the same asymptotic with famous logarithmic
correction~\cite{giuliani_lifetime_1982} as for HEG 2D,
Eq.~\eqref{eq:tau:2d}~\cite{li_finite_2013}. Polini and Vignale provided a very
pedagogical derivation of this fact~\cite{polini_quasiparticle_2014}, however, exchange
contributions were not included. Our simulations show that they are indeed small for MLG
and BLG systems, Fig.~\ref{fig:Gm}. One interesting conclusion of
Ref.~\cite{polini_quasiparticle_2014} is that the scatterings are dominated by the
``collinear scattering singularity'', that is, the momenta of electrons involved in the
scattering are mostly parallel to each other. We find it interesting because of its
apparent similarities with the scattering processes shaping $\Sigma_{2x}$, see our
analysis in Sec.~\ref{sec:sgm2x:heg}. Another interesting conclusion of the analytical
formula is that the life-time is independent of the dielectric constant. While our
numerics shows that its only approximately true for full-fledged calculations using
Eq.~\eqref{eq:tau:def}, an illustration on why this is the case for the on-shell
$\tau^{-1}$ is provided in Fig.~\ref{fig:Im:SGM:kp}. There, we plot the imaginary
self-energy part for $k=1.6\kF$ and three different dielectric constants,
$\kappa=0.5,\,1.0,\,2.5$ resolving plasmonic and $p$-$h$ contributions. Approaching the
``on-shell'' frequency marked as a vertical dashed line, the curves essentially fall on
top of each other. The figure also illustrates the difficulties of the numerical
determination of the asymptotic prefactors in Eq.~\eqref{eq:tau:2d} (for MLG they also
depend on the momentum cut-off).

%=======================================================================================
%                                  Sec. VI. CONCLUSIONS         ////////////////////////
%=======================================================================================
\section{Conclusions\label{sec:conclusions}}
Hedin's set of functional equations allows us to expand the electron self-energy in terms
of dressed $G$ and $W$. The first term of such an expansion\,---\,the $GW$
approximation\,---\,has been successfully applied to many systems. However, there are
cases when this approximation is insufficient and higher-order terms need to be taken into
account. One such approximation has been derived in our previous work starting from the
first- and second-order
SE~\cite{stefanucci_diagrammatic_2014}. $\Sigma_{aa}+[\Sigma_{cc}+\Sigma_{c\bar{c}}]+\Sigma_{a\bar{a}}$
describes three distinct scattering processes in many-body systems, comprises all the
first- and second-order terms and a subset of third and fourth order terms, and,
crucially, has the PSD property~\cite{pavlyukh_vertex_2016}. In this work focused on
$\Sigma_{a\bar{a}}$, relevant for small energy transfers, and evaluated it in the
quasiparticle approximation for the electron GF and RPA for the screened interaction. We
found that screening is important and must be determined consistently. For inconsistent
screening, unphysical singularities have been observed in $\Sigma_{2x}$, which is the bare
Coulomb limit of $\Sigma_{a\bar{a}}$. Nonetheless, $\Sigma_{2x}$ provides important
corrections to the total energy in full agreement with analytic results of Onsager
\emph{et al.}~\cite{onsager_integrals_1966}.

We conducted a comprehensive investigation of the impact of $\Sigma_{a\bar{a}}$ on
quasiparticle properties of the homogeneous electron gas in 2D and 3D, and of the mono-
and bilayer graphene. The quasiparticle renormalization factor $Z_{qp}(\kF)$, the
effective mass $m^{*}$, the Fermi velocity $\vF^{*}$, and the quasiparticle life-time
$\tau(k)$ have been computed for a range of interaction strengths controlled by the
density $r_s$ or the dielectric function $\kappa$. In the weakly correlated limit
($r_s\ll1$ or $\kappa\gg1$), we compared with asymptotic expansions, and in the correlated
regime with results of other theories such as quantum Monte Carlo and perturbative
calculations including local field factors.

It is known that exchange processes encoded in $\Sigma_{a\bar{a}}$ reduce the
quasiparticle scattering rate. This has been shown in the asymptotic limit
$\w\rightarrow\infty$ by Vogt \emph{et al.}~\cite{vogt_spectral_2004} and in the vicinity
of $\kF$ by Qian and Vignale~\cite{qian_lifetime_2005}. Besides confirming this finding
using a completely different methodology, we also observed an appreciable effect of
$\Sigma_{a\bar{a}}$ on the effective mass $m^{*}$ and quasiparticle strength $Z_{qp}$ in
3D HEG. The effect is smaller in 2D, especially for graphene systems, as anticipated.

Although we have focused on one particular scattering process the PSD diagrammatic
construction is versatile and applicable to realistic multiband systems. Furthermore, the
PSD diagrams remain PSD upon replacement of the zero-temperature Green’s function with the
finite-temperature~\cite{kas_finite_2017} or any excited-state Green’s function.  This
latter possibility opens the way toward the systematic inclusion of vertex corrections in
the spectral function of systems in a (quasi) steady-state. Investigations in the field
of, e.g., molecular transport and time-resolved (tr) and angle-resolved photoemission
spectroscopy (ARPES) are therefore foreseeable in the near future. The spectral function
is indeed a key quantity to determine the conductance of atomic-scale junctions, and MBPT
calculations have so far been limited to the
$GW$~\cite{thygesen_nonequilibrium_2007,spataru_gw_2009,neaton_renormalization_2006,
  myohanen_many-body_2008,myohanen_kadanoff-baym_2009} and
second-Born~\cite{myohanen_many-body_2008,myohanen_kadanoff-baym_2009} approximation.
Similarly, the tr-ARPES signal is related to the transient spectral
function~\cite{freericks_theoretical_2009,perfetto_first-principles_2016} which in
semiconductor or insulators can be evaluated using a steady-state approximation (provided
that the carrier relaxation time is much longer than the probe pulse).  In this
context~\cite{hyrkas_diagrammatic_2019} the PSD diagrammatic construction may provide a
powerful tool in the field of light-induced exciton fluids, whose incoherent plasma
phase~\cite{semkat_ionization_2009,perfetto_first-principles_2016,steinhoff_exciton_2017}
and coherent condensed
phase~\cite{rustagi_photoemission_2018,rustagi_coherent_2019,perfetto_pump-driven_2019,
  christiansen_theory_2019,perfetto_time-resolved_2020,hanai_non-equilibrium_2016,becker_projector-based_2019}
are currently under intense investigations.

\begin{acknowledgments}
 We thank A.-M.\,Uimonen for useful discussions. The work has been performed under the
 Project HPC-EUROPA3 (INFRAIA-2016-1-730897), with the support of the EC Research
 Innovation Action under the H2020 Programme; in particular, Y.P. gratefully acknowledges
 the computer resources and technical support provided the CSC-IT Center for Science
 (Espoo, Finland). Y.P. acknowledges support of Deutsche Forschungsgemeinschaft (DFG),
 Collaborative Research Centre SFB/TRR 173 ``Spin+X''. G.S. acknowledges funding from MIUR
 PRIN Grant No. 20173B72NB and from INFN17\textunderscore nemesys project. R.vL. likes to thank the
 Academy of Finland for support under grant no. 317139.
\end{acknowledgments}
\appendix
\section{Hilbert transform and spectral functions\label{sec:app:hilbert}}
Hilbert transform is an important part of our numerical procedure. We define
\begin{align}
  H\mleft[x\mright](t)&=\frac1\pi\mP\int_{-\infty}^{\infty} \!\!\dd\tau\,\frac{x(\tau)}{t-\tau}.
  \label{eq:Hilbert}
\end{align}
It has the properties $H\mleft[H[x]\mright](t)=-x(t)$,
$H^{-1}\mleft[x\mright](t)=-H\mleft[x\mright](t)$, and is computed using FFT. In
particular we need the relation between the real part of the correlation self-energy and
the rate function $\Gamma(k,\w)$, Eq.~\eqref{eq:rate}, which in our approach is computed
by the Monte-Carlo method,
\begin{align}
  \re\SgRc(k,\w)&=\tfrac12 H[\Gamma(k)](\w),\\
  \tfrac12 \Gamma(k,\w)&=-\im\SgRc(k,\w).
\end{align}
There are following possibilities to obtain positive spectral functions starting from the
second-order self-energy:
\begin{subequations}
\begin{align}
  \mp i \Sigma_{aa}^\lessgtr(k,\omega)&\ge0,\\
  \mp i\left(\Sigma_{cc}^\lessgtr+\Sigma_{c\bar{c}}^\lessgtr\right)(k,\omega)&\ge0,\\
  \mp i\left(\Sigma_{aa}^\lessgtr+\Sigma_{a\bar{a}}^\lessgtr\right)(k,\omega)&\ge0.
\end{align}
\label{eq:subs:sgm:psd}
\end{subequations}
Consequently, the sum of all contributions given by Eq.~\eqref{eq:sgm_abc} is also PSD. By
using the method from our earlier work~\cite{stefanucci_diagrammatic_2014}, these results
can also be generalized to any dressed GFs that possess a positive spectral function.
\section{Equilibrium propagators\label{sec:app:prop}}
We define the bare electron propagators as averages of the field operators in the
Heisenberg picture over the non-interacting state
\begin{align*}
  g^<(1,2)&=\ii \langle\hat\psi_H^\dagger(2)\hat\psi_H(1)\rangle_0,&
  g^>(1,2)&=-\ii \langle\hat\psi_H(1)\hat\psi_H^\dagger(2)\rangle_0,
\end{align*}
fulfilling the symmetry relations
\begin{align}
  \ii g^{\lessgtr}(1,2)&=\left(\ii g^{\lessgtr}(2,1)\right)^*.
\end{align}
Analogically, the density-density correlators are defined with respect to the interacting
ground state
\begin{align*}
\chi^{>} (1,2)&=-\ii \bra \Delta\hat{n}_{H}(1) \Delta\hat{n}_{H}(2)\ket,  \\
\chi^{<} (1,2)&=-\ii \bra  \Delta\hat{n}_{H}(2) \Delta\hat{n}_{H}(1) \ket, 
\end{align*}
with the density deviation $\Delta\hat{n}_H (1) =\hat{n}_H (1) -\bra \hat{n}_H(1) \ket $.
They fulfill the symmetries
\begin{align}
  \ii \chi^{\lessgtr} (1,2)&=\left(\ii \chi^{\lessgtr}(2,1)\right)^*.
\end{align}
Because the screened interaction is directly related to $\chi$
\begin{align}
  W(1,2)&=v(1,2)+\iint\dd(3,4)\,v(1,3)\chi(3,4)v(4,3),
\end{align}
all the symmetry and analytic properties also hold for $W$.

For homogeneous systems the momentum-energy representation is useful, which we formulate
here in Fermi units ($\kF$, $\eF$). The Kubo-Martin-Schwinger conditions allow us to write
the lesser/greater propagators in terms of the retarded ones,
\begin{subequations}
\begin{align}
  g^<(x,\zeta)&=-2\ii \nF(\zeta)\im \gR(x,\zeta),\\
  g^>(x,\zeta)&= -2\ii (\nF(\zeta)-1)\im \gR(x,\zeta);\\
  W_0^<(y,\xi)&=+2\ii \nB(\xi)\im \WnR(y,\xi),\\
  W_0^>(y,\xi)&=+2\ii (\nB(\xi)+1)\im \WnR(y,\xi).
\end{align}
\label{eq:prop:bare:def}
\end{subequations}
$n_\text{F/B}$ are the Fermi/Bose distribution functions, which at zero temperature reduce
to simple step-functions
\begin{subequations}
\begin{align}
  \nF(\zeta)&=\theta(1-\zeta),&\nbF(\zeta)=1-\nF(\zeta);\\
  \nB(\xi)&=-\theta(-\xi)=\theta(\xi)-1,&\nB(\xi)+1=\theta(\xi).
\end{align}
\label{eq:nF:nB}
\end{subequations}
For the bare propagators we furthermore have
\begin{align}
  \gR(x,\zeta)&=\frac{1}{\zeta-\epsilon(x)+\ii \eta},
  \label{eq:gR:bare:def}
\end{align}
and we use a spectral representation of the screened interaction
\begin{align}
  \WnR(y,\xi)&
  =v(y)+\int_{0}^\infty\!\dd\w \frac{2\w\, C(y,\xi)}{(\xi+\ii\eta)^2-\w^2},
  \label{eq:W:R}
\end{align}
where $v(y)$ is the bare Coulomb interaction. It fulfills the symmetry property
\begin{align}
  \left[\WnR(y,\xi)\right]^*&=\WnR(y,-\xi).
\end{align}
Comparing it with the Hilbert transform of the inverse dielectric function
\begin{align}
  \frac1{\eR(y,\xi)}&
  =1-\int_{0}^\infty\!\frac{\dd\w}{\pi}\im \mleft[\frac1{\eR(y,\w)}\mright] \frac{2\w}{(\xi+\ii\eta)^2-\w^2},
  \label{eq:inv:eps:2}
\end{align}
we have for the spectral function of continuous spectrum
\begin{align}
  C(y,\w)&=v(y)\im \mleft[-\frac{1}{\pi}\frac1{\eR(y,\w)}\mright],
  \label{eq:C:ph:def}
\end{align}
and for plasmons $C(y,\w)=C(y)\delta\mleft(\w-\Omega(y)\mright)$ with
\begin{align}
  C(y)&=v(y)\left[\left.\frac{\partial \re\eR(y,\xi)}{\partial \xi}\right|_{\xi=\Omega(y)} \right]^{-1}.
  \label{eq:C:pl:def}
\end{align}
The time-ordered ($\WnT\equiv W_0^{--}$) and the anti-time-ordered
($\WnAT\equiv W_0^{++}$ ) screened interactions read
\begin{align}
   \WnT(y,\xi)& =v(y)+\int_{0}^\infty\!\dd\w \frac{2\w\, C(y,\w)}{\xi^2-(\w-\ii\eta)^2},
   \label{eq:W:C:A}\\
   \WnAT(y,\xi)& =-\left[\WnT(y,\xi)\right]^*,
\end{align}
with $\WnT(y,-\xi)=\WnT(y,\xi)$.
\section{Polarizability  of MLG\label{app:mlg}}
The dynamical polarization of graphene at finite doping has been computed by Hwang and Das
Sarma~\cite{hwang_dielectric_2007} and by Wunsch \emph{et
  al.}~\cite{wunsch_dynamical_2006}. We present here for completeness the functions $G_r$,
\begin{align}
G_r(y,\xi)&=\begin{dcases*}
    0&1A\\
    \pi+G_<(z_1)&2A\\
    \pi+G_<(z_1)+G_<(z_2)&3A\\
    -G_>(z_2)+G_>(-z_1)&1B\\
    -G_>(z_2)&2B\\
    -G_>(z_2)+G_>(z_1)&3B
\end{dcases*},
\end{align}
and $G_i$,
\begin{align}
G_i(y,\xi)&=
  \begin{dcases*}
    G_>(z_2)-G_>(-z_1)&1A\\
    G_>(z_2)&2A\\
    0&3A\\
    0&1B\\
    \pi+G_<(z_1)&2B\\
    \pi&3B
  \end{dcases*}.
\end{align}
that define the dielectric function in Eqs.~(\ref{eq:eps:mlg:re},\ref{eq:eps:mlg:im}) with
\begin{align}
  z_1&=\frac{\xi-2}{y},&
  z_2&=\frac{\xi+2}{y},
\end{align}
and
\begin{align}
  G_>(z)&=z\sqrt{z^2-1}-\mathrm{arccosh}(z),\\
  G_<(z)&=z\sqrt{1-z^2}-\arccos(z).
\end{align}
Notice that A-domains are for $\xi<y$ and $B$-domains are for $\xi>y$ as shown in
Fig.~\ref{fig:chi:dom}.
\section{Polarizability of BLG\label{app:blg}}
We start by defining the four critical lines
\begin{align*}
  r_1&=y^2+2 y+\xi+2;&r_2&=y^2+\xi;\\
  r_3&=y^2-2 y+\xi+2;&r_4&=\frac{y^2}{2}+\xi.
\end{align*}
Furthermore, we introduce some auxiliary functions:
\begin{align} 
  q_1&=\frac{y^2-2\xi}{4\xi}\log2+\frac{r_4}{2\xi}\log|r_4|-\frac{r_2}{2\xi}\log|r_2|\thf(y-1)\nn\\
  &\qquad+\left(\frac{r_2}{4\xi}\log|1+\xi|+\frac{y^2}{4\xi}\log|y^2|\right)\mathrm{sign}(y-1),\\
  q_2&=\left[\frac{|r_2|}{4\xi}\log\left|\frac{(2+\xi)\sqrt{r_2 r_2}+\xi\sqrt{r_1r_3}}{(2+\xi)\sqrt{r_2 r_2}-\xi\sqrt{r_1r_3}}\right|\right.\nn\\
    &\quad\left.-\frac14\log\left|\frac{\sqrt{r_1r_3}+2+\xi}{\sqrt{r_1r_3}-(2+\xi)}\right|\right]\left(\thf(-r_1)-\thf(r_3)\right),
\end{align}
\begin{align} 
  p_1&=\left[\frac{r_2}{2\xi}\arctan\mleft(\frac{(2+\xi)r_2}{\xi\sqrt{-r_1r_3}}\mright)\right.\nn\\
    &\qquad\left.+\frac12\arccos\mleft(\frac{2+\xi}{y\sqrt{-2r_4}}\mright)\right]\thf(-r_1r_3),\\
  p_2&=\frac{\pi r_2}{4\xi}\mathrm{sign}(r_2)\thf(-r_1r_3)+\frac{\pi r_4}{\xi}\thf(y-2)\thf(-r_4)\thf(r_3)\nn\\
  &\qquad-\frac{\pi y^2}{2\xi}\thf(-r_1).
\end{align}
With the help of these definitions
\begin{align}
  \re \Pi_2(y,\xi)&=q_1+q_2;&
  \im \Pi_2(y,\xi)&=p_1+p_2.
\end{align}

\section{Solution of the Dyson equation\label{app:dyson}}
Let us recapitulate possible approaches to the solution of the Dyson equation
\begin{align}
  G(k,\w)&=g(k,\w)+g(k,\w)\Sigma(k,\w)G(k,\w)
  \label{eq:dyson}
\end{align}
following Ref.~\cite{stefanucci_nonequilibrium_2013}. In the preceding sections the
self-energy is computed using bare propagators~\eqref{eq:prop:bare:def},
$\Sigma=\Sigma[g,W_0]$. This approach has an inherent problem that the $k=\kF$ state is
\emph{no longer a sharp quasiparticle state}. We still can improve the \emph{one-shot}
calculations by applying some rigid shift $\Delta \mu$ to all poles,
\begin{align}
  G_0(k,\w)&=\frac{1}{\w-\epsilon_0(k)-\Delta \mu+\ii\eta}=g(k,\w-\Delta\mu).
  \label{eq:g:0:def}
\end{align}
The respective self-energy then reads
\begin{align}
  \Sigma[G_0,W_0](k,\w)&=\Sigma[g,W_0](k,\w-\Delta \mu),
\end{align}
allowing to rewrite the quasiparticle approximation for the Dyson's equation
\begin{align}
  \epsilon(k)&=\epsilon_0(k)+\Sigma[g,W_0](k,\epsilon(k)-\Delta\mu),\label{eq:Dyson:1}\\
  \tilde{\epsilon}(k)&=\epsilon_0(k)-\Delta\mu+\Sigma[g,W_0](k,\tilde{\epsilon}(k)),
\end{align}
Thus, the quasiparticle approximation for $G$ reads
\begin{align}
  G(k,\w)&=\frac{1}{\w-\tilde{\epsilon}(k)}.
\end{align}
Now we demand that the solution of the Dyson equation at $k=\kF$ takes the form
\begin{align}
  G(\kF,\w)&=\frac{1}{\w-\mu+\ii\eta}=G_0(\kF,\w).\label{eq:qp:mu}
\end{align}
and coincides with the improved propagator $G_0$~\eqref{eq:g:0:def} representing a sharp
quasiparticle peak at the chemical potential. The consistency condition~\eqref{eq:qp:mu}
provides the interpretation of $\Delta\mu$ as the correlation shift of the chemical
potential
\begin{align}
  \Delta \mu&=\mu-\eF,\label{eq:dmu:def}
\end{align}
and allows to determine it. To this end, we insert Eq.~\eqref{eq:qp:mu} into
Eq.~\eqref{eq:Dyson:1} leading to
\begin{align}
  \mu&=\eF+\re\Sigma[G_0,W_0](\kF,\mu).
  \label{eq:mu}
\end{align}
This point and the connection of $\mu$ to the \emph{total energy per electron} is
explained in Ref.~\cite{hedin_effects_1970} (p.\,82). Combining Eq.~\eqref{eq:dmu:def}
with Eq.~\eqref{eq:mu} we obtain
\begin{align}
  \Delta\mu&=\re\Sigma[g,W_0](\kF,\eF).\label{eq:dmu:comp}
\end{align}
Thus, $\Delta\mu$ is expressed solely in terms of the self-energy for $k=\kF$ and
$\w=\eF$. In the case of MLG and BLG having two bands, one additionally sets the band
index $s$ consistent with the doping (typically the chemical potential is \emph{above} the
Dirac point implying $s=+1$).

%\bibliography{MyLibrary}
%

\end{document}